\documentclass[letter]{aa}

\usepackage[varg]{txfonts}
\usepackage{graphicx}
\usepackage{natbib}
\usepackage{xspace}
\usepackage[colorlinks,citecolor=blue,linkcolor=blue,urlcolor=blue]{hyperref} 

\newcommand{\Msun}{\ensuremath{M_{\odot}}\xspace}

\newcommand{\SE}{SESNe\xspace}
\newcommand{\SNI}{SN\,Ib/c\xspace}
\newcommand{\SNIIb}{SN\,IIb\xspace}

\definecolor{amethyst}{rgb}{0.6,0.4,0.8}

\def\posydon{\texttt{POSYDON}~}

\begin{document}

\title{Revisiting the explodability of single massive star progenitors of stripped-envelope supernovae}


\author{E.~Zapartas \inst{1}
\and
M.~Renzo\inst{2,3}
\and
T.\,Fragos\inst{1}
\and
A.\,Dotter\inst{4}
\and
J.\,J.\,Andrews\inst{4}
\and
S.\,S.\,Bavera\inst{1}
\and
S.\,Coughlin\inst{4}
\and
D.\,Misra\inst{1}
\and
K.\,Kovlakas\inst{1}
\and
J.\,Rom\'an-Garza\inst{1}
\and
J.\,G.\,Serra\inst{4}
\and
Y.\,Qin\inst{5}
\and
K.\,A.\,Rocha\inst{4}
\and
N.\,H.\,Tran\inst{6}
\and
Z.P.\,Xing\inst{1}
}

\institute{Département d’Astronomie, Université de Genève, Chemin Pegasi 51, CH-1290 Versoix, Switzerland.
\and
Department of Physics, Columbia University, New York, NY 10027, USA
\and
Center for Computational Astrophysics, Flatiron  Institute, 162 5th Ave, New York, NY 10010, USA
\and
Center for Interdisciplinary Exploration and Research in Astrophysics (CIERA) and Department of Physics and Astronomy, Northwestern University, 1800 Sherman Avenue, Evanston, IL 60201, USA
\and
Department of Physics, Anhui Normal University, Wuhu city, Anhui Province, 241000, P.R. China
\and
DARK, Niels Bohr Institute, University of Copenhagen, Jagtvej 128, DK-2200 Copenhagen, Denmark\\
\email{ezapartas@gmail.com}
}


\abstract{ Stripped-envelope supernovae (Types IIb, Ib, and Ic) that show
  little or no hydrogen comprise roughly one-third of the
    observed explosions of massive stars. Their origin and the
  evolution of their progenitors are not yet fully understood. Very
  massive single stars stripped by their own winds
  ($\gtrsim 25-30 \Msun$ at solar metallicity) are considered viable
  progenitors of these events. However, recent 1D core-collapse
  simulations show that some massive stars may collapse directly into
  black holes after a failed explosion, with a weak or no visible
  transient. In this letter, we estimate the effect of direct collapse
  into a black hole on the rates of stripped-envelope supernovae that
  arise from single stars. For this, we compute single-star MESA
  models at solar metallicity and map their final state to their
  core-collapse outcome following prescriptions commonly used in
  population synthesis.  According to our models, no single stars that
  have lost their entire hydrogen-rich envelope are able to explode,
  and only a fraction of progenitors left with a thin hydrogen envelope
do (IIb progenitor candidates), unless we use a 
    prescription that takes the effect of turbulence into account  or invoke increased wind mass-loss rates.
  This result increases the existing tension between the single-star 
 paradigm to explain most 
  stripped-envelope supernovae and their observed rates
  and properties.
  At face value, our results  point toward an even higher contribution of binary progenitors to stripped-envelope supernovae. Alternatively, they may suggest  inconsistencies in the common practice of mapping different stellar models  to core-collapse outcomes and/or higher overall mass loss in massive stars.
  }


\date{Received 09/06/2021; accepted 23/09/2021}

\titlerunning{The explodability of single stripped-envelope supernova progenitors}

\maketitle

\section{Introduction}\label{sec:intro}

Core-collapse supernovae (SNe) mark the explosive death of massive
stars. A subset of them show no observational features of hydrogen  (H) in
their spectra.  This indicates that their progenitors had been
(almost) fully stripped of their H-rich envelopes before
explosion. These so-called stripped-envelope SNe (SESNe) are
observationally classified as Type Ib or Ic (\SNI) when they are fully
stripped of H, with Ic also lacking signatures of helium (He)
in their spectra. Type IIb SNe (\SNIIb) are a transitional  subclass of SESNe that show H
lines only for a few days after the explosion, which subsequently disappear \citep[for a review of SN classification, see][]{Filippenko1997}. 

One mechanism proposed for the production of SNe Ib/c
is the stripping of the hydrogen-rich envelope of massive single-star progenitors due
to their own strong mass loss \citep[e.g.,][]{Maeder+1982,Heger+2003,Georgy+2012}. 
These stripped stars might appear as Wolf-Rayet (WR) stars with emission lines, or not, depending on their wind mass-loss rates \citep[e.g.,][]{Crowther2007, Gotberg+2018}.  
The loss of the outer layers is the outcome of many possible mass-loss
mechanisms throughout the stellar lifetime \citep[for a review, see][]{Smith2014}.

Recently, there has been an ongoing investigation into whether all
massive star progenitors of a core-collapse event produce observable
transients.  Self-consistent simulations that model the core-collapse
process, within the neutrino-driven explosion paradigm, have suggested
the possibility that the SN shock might stall and be reverted
by the infalling material, failing to unbind the outer layers of the
star
\citep[][]{Herant+1994,Burrows+1995,Fryer+Heger2000,Janka2013,
    Muller2019, Vartanyan+2021}. In this case, in the
  absence of an extended H-rich envelope \citep[e.g.,][]{quataert:19, antoni:21},
the expected outcome
is the implosion of the collapsing star into a black hole (BH), with only a weak transient
\citep[e.g.,][]{Nadezhin1980, 
Fernandez+2018, ivanov:21} or possibly no transient at all. 
 There have been observational efforts to identify the disappearance
 of evolved stars without a SN, and thus far there are two potential
 candidates of red supergiant stars that have been reported missing
 \citep{Gerke+2015, Adams+2017a, Adams+2017b, Neustadt+2021}.

 The uncertainty of the outcome of the collapse process is an
   important general limiting factor in the interpretation of
   gravitational wave detections
   \citep[e.g.,][]{Belczynski+2016,Bavera+2020}, X-ray binary
   kinematics \citep[e.g.,][]{Atri+2019}, and SN statistics, all of
   which are seeing rapid observational advancements with LIGO/Virgo \citep[e.g.,][]{Abbott+2021}, \emph{Gaia}
   \citep{Gaia-Collaboration2018}, and large transient surveys
   \citep[e.g.,][]{Perley+2020}, respectively. More specifically,
   robotic surveys such as the Vera C. Rubin Observatory’s Legacy
   Survey of Space and Time \citep{ivezi:2019} will provide large
   samples of transients, including \SE; however, \emph{ab initio}
   simulations of stellar explosions are still sparse, and theoretical
   rate calculations need to rely on simpler parametric core-collapse
   models \citep[e.g.,][]{OConnor+2011, Fryer+2012,
     Sukhbold+2014,Sukhbold+2016, Ertl+2016, Ebinger+2019, Ertl+2020,
     Patton+2020, Couch+2020}. Here, we investigate the consistency
   between the single-star progenitors self-stripped by their own
   winds to produce \SE and a wide range of state-of-the-art
   parametric explosion models.

 The stellar core structure determines the collapse, explosion,
   and, ultimately, the occurrence of a bright SN transient. Only by
 parametrizing the complex pre-SN core structure and the
 multidimensional collapse and explosion dynamics is it possible to
 explore a wide range of progenitors. One-dimensional (1D) parametric core-collapse models
 predict a non-monotonic behavior of the properties of this inner core
 with initial mass, resulting in ``islands'' of successful explosions  in the mass parameter space, which are surrounded  by regions of BH formation via
 direct collapse. 
 
The observed relative rate of \SE to all core-collapse SNe is
   $\sim{}1/3$ \citep{Smith+2011, Eldridge+2013, Shivvers+2019}. This
   ratio is not easily explained if only single massive
   stars are considered: A large initial mass is required for them to lose all their
   H-rich material, which makes them disfavored by the initial mass
   function \citep[e.g.,][]{Smith+2011} and generally makes it harder for them to
   explode successfully.  If even a fraction of them do not produce a
   transient at collapse, it would increase the significance of the
   inconsistency between the observed rates and the assumption that most \SE come from single
   massive progenitors. 
Our rate of  massive single
   stars successfully exploding as \SE provides an independent constraint to
   be taken into account in the interpretation of 
   observational data and of the outcome of population synthesis and
   parametric SN explosion simulations\footnote{Very recently, \citet{Patton+2021} carried out
   a study in a similar direction, focusing on the compact object
   remnant masses produced by these prescriptions.}.

\section{Method}\label{sec:method}

In this study we estimated the rate of successful hydrogen-poor SN
explosions from isolated stars using the software framework \posydon 
(Fragos et
al. 2021, in prep.).\footnote{\url{https://posydon.org/}} Our default
results are based on a grid of 96 single-star models at solar
metallicity \citep[$Z_{\odot}=0.0142$;][]{Asplund+2009}.
The grid
  spans initial masses from 10 to 120 \Msun, with a spacing of 1 \Msun between 10-25
  \Msun, 0.5\Msun between 25-55\Msun to more accurately capture the
  transition to partially and fully stripped stars, and
  logarithmically spaced above 55\Msun. The stellar models evolved using the MESA 1D stellar evolution code \citep[][version
11701]{Paxton+2011,Paxton+2013, Paxton+2015, Paxton+2018, Paxton+2019} from zero-age main sequence up to carbon core depletion. At that evolutionary stage the star has less than a few years until its core collapses, and its surface chemical profile is not expected to vary significantly until the end. This would change in the case of  possible episodic mass loss in the remaining stages, which we do not account for but discuss in Sect.~\ref{sec:conclusions}. 
We briefly summarize here some of the key
physical assumptions as well as certain variations of them that we explore. 

In our default model we assumed standard ``Dutch'' wind mass-loss prescriptions in various
evolutionary phases, following \citet{Vink+2000} for main-sequence stars and switching to \citet{de-Jager+1988} whenever an evolved
star expands and its effective temperature drops below $ 10^4 K$. We
assumed a mass-loss rate, following \citet{Nugis+2000}, when the surface
 H mass fraction drops below $0.4$. 
We also experimented with scaling up wind mass loss by $1.5$.

We adopted the Ledoux criterion for convective instability, with a mixing length
parameter of $\alpha_{\mathrm{MLT}} {=} 1.93$ \citep{Choi+2016}. In our default model we assumed an exponential
convective overshooting only during the main-sequence phase, with
$f_{\mathrm{ov}}= 4.15 \times 10^{-2} $ and $f_{0,\mathrm{ov}}= 8 \times 10^{-3} $, calibrated following \citet{Claret+Torres2017} to produce
main-sequence evolutionary tracks similar to the step overshooting of  \citet{Brott+2011a}.
We also studied a population with models where there is the same over-
and undershooting in shell burning regions, as well as one with a lower core overshooting of $f_{\mathrm{ov}}= 1.6 \times 10^{-2} $  during main sequence \citep{Choi+2016}. All the above assumptions regarding overshooting lead to
lower helium and carbon-oxygen core masses.
We assumed zero stellar rotation in our default simulation, but we also 
explored a variation where we assumed that stars are born at 40\% of
their critical rotation, 
allowing for rotational chemical mixing
\citep{Chaboyer+1992,Heger+2000} and 
angular momentum transport \citep{Spruit2002}. 

We defined as possible progenitors of SNe Ib or SNe Ic %
(SN Ib/c  ``candidates'')
all stars that expel their H-rich envelopes prior to carbon depletion ($M_{\rm env} = 0 $)
 and are left only with H-deficient ``helium''
stellar cores, with a boundary defined by the low hydrogen mass fraction of $X_{\rm H}<10^{-2}$. 
We refrained from differentiating between SN Ib and SN Ic progenitors due to the
uncertainty in the expected surface structure and abundances that lead
to each subclass \citep[][although see also \citealt{Modjaz+2016}]{Dessart+2012}.
We also assumed that a successfully exploding star with a thin H-rich layer, $0  < M_{\rm env} \leq 0.5 \Msun$, will produce a SN IIb \citep[e.g.,][]{ Maund+2004, van-Dyk+2011, Bersten+2012}.

To determine whether a star will produce a successful explosion, we mapped its structure at carbon depletion  to its  ``explodability'' properties, following various prescriptions available in the literature from \citet{Fryer+2012}, \citet{Sukhbold+2016}, \citet{Patton+2020}, and \citet{Couch+2020}. 
%
\citet{Fryer+2012} introduced two prescriptions, ``rapid'' and
``delayed'' (henceforth F12+rapid and F12+delayed, respectively), to
determine the final compact remnant type and mass, based on their
final carbon-oxygen core mass, $M_{\rm CO}$. The main difference between the two prescriptions for our study is that  F12+rapid produces a possibly physical	 mass gap between neutron stars and BHs in the $2-5$ \Msun mass range. 

\begin{figure}[t]\center
\includegraphics[width=0.49\textwidth]{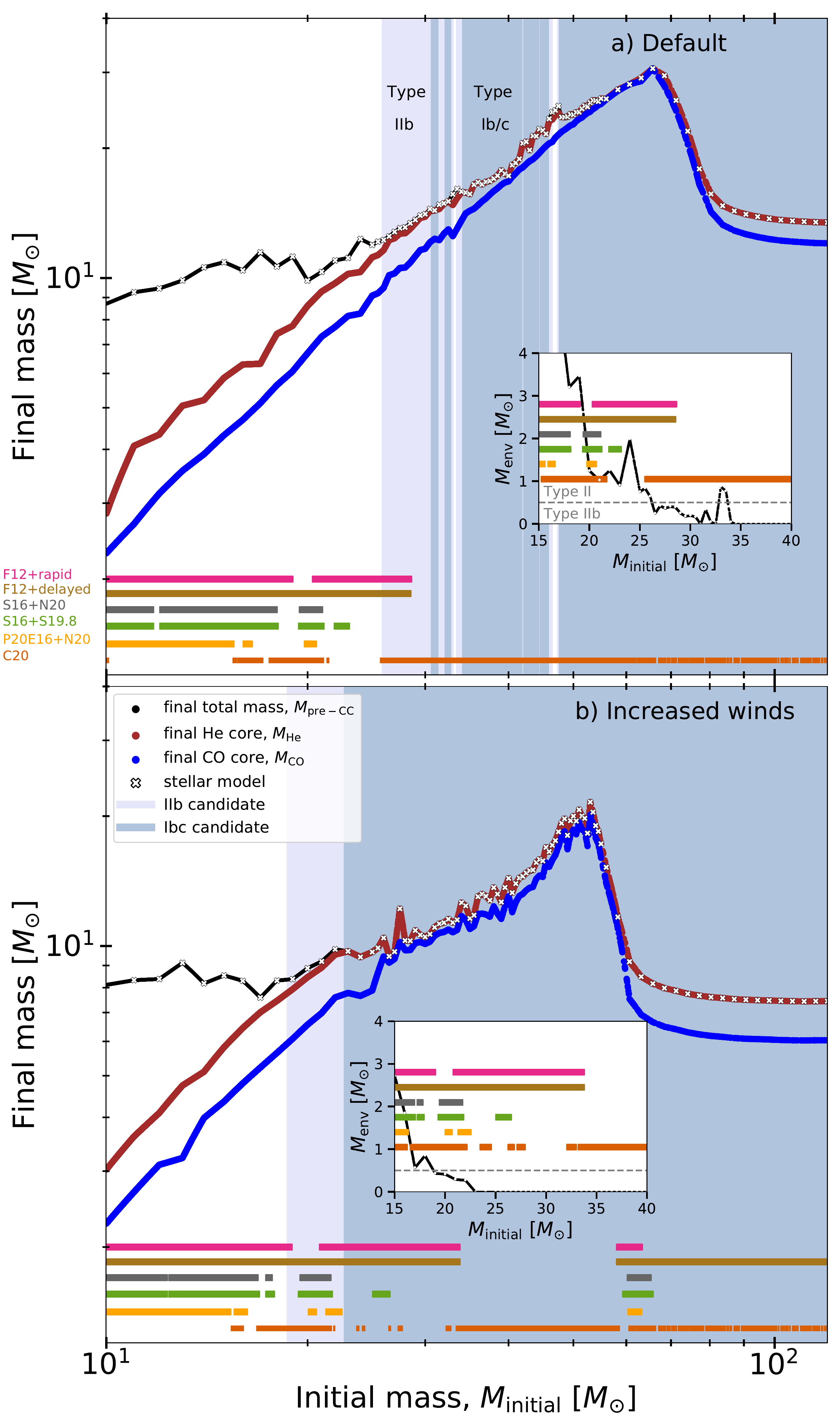}
\caption{
  Final total mass $(M_{\rm final}$; black), helium core mass ($M_{\rm He}$; maroon), and carbon-oxygen core mass
($M_{\rm CO}$; blue) as a function of initial
  mass ($M_{\rm initial}$) for our default grid (top) and our grid with winds boosted by a factor of 1.5 (bottom). We also depict the stellar models (white crosses) that we interpolate between. In the background, we shade the mass ranges for final
  stellar structures that could give rise to SNe IIb if they explode
  at collapse (with $0 < M_{\rm env} \leq 0.5 \Msun$; light blue
  shading) and WR stars that may produce SNe Ib/c (with $M_{\rm env} = 0$, blue shading). 
  We also mark the models that lead to successful
  explosions, according to the FR12-rapid, FR12-delayed, S16-N20, S16-S19.8, P20E16+N20, and C20 SN  engines
  (horizontal  magenta, brown, gray, light green, light orange, and dark orange bands of arbitrary y value). In the inset plots we show the final envelope mass ($M_{\rm env}$) for models with initial masses of $15-40\Msun$.  
}
\label{fig:Mfinal}
\end{figure}

The outcome of the collapse of \citet{Sukhbold+2016} pre-SN models has been calibrated against the well-studied SN 1987A progenitor.
From the different SN engine calibrations available, we followed their most optimistic options for successful explosions of  ``N20'' and  ``S19.8'' (S16+N20 and S16+S19.8, respectively). 
In our study, following \citet{Roman-Garza+2021},
for every star 
in our population 
we assumed the same explodability outcome as the \citet{Sukhbold+2016} model with the closest final helium core mass, $M_{\rm He}$.  This is referred to as the ``hybrid'' method in \citealt{Patton+2021}.
For simplicity, we neglected the rare occurrence of any fallback-powered explosions during BH formation.

\citet[][henceforth C20]{Couch+2020} included the effect of turbulence
  during the collapse simulated in 1D \citep[see also][]{Mabanta+Murphy2018,Mabanta+2019}. They applied their explosion
  model to the progenitors from \citet{Sukhbold+2016}, so we followed
  the same hybrid approach as in the S16+N20 and S16+S19.8 prescriptionss, assuming
  an explodability outcome based on the final helium core mass. We used
  a value of $\alpha_{\Lambda} = 1.2$ for the free parameter
  that describes the effect of turbulence, consistent with their favored 
  range.  As already noted by C20, the explosion
  outcomes are radically different from the results of
  \citet{Sukhbold+2016} and \citet{Ertl+2016}.

In the \citet{Patton+2020} prescription, the carbon abundance and core mass at carbon ignition 
determine the explodability of the pre-SN core. The carbon abundance
is sensitive to the assumed  $^{12}{\rm C}(\alpha,\gamma)^{16}{\rm O}$
reaction rate \citep[in our case, from][]{Cyburt+2010}\footnote{\url{https://reaclib.jinaweb.org/index.php}}, which is notoriously uncertain
\citep[e.g.,][]{Holt+2019, Farmer+2020}. \citet{Patton+2020}
  study CO core masses up to 10 \Msun; they argue that with higher core
  masses it is even more difficult to explode and that the formation of a BH is most likely \citep{Sukhbold+2016,Ertl+2020}. We thus followed the same
  assumption here. We could then map our models to the explodability
parametrization of \citet{Ertl+2016},
 arguing that it captures  the results from 3D simulations 
more accurately compared to the compactness parametrization of \citet{OConnor+2011}. We mapped to the N20 calibration option of \citet{Ertl+2016} to determine the outcome of the collapse process (SN engine  P20E16+N20), although we verified that the results are similar for the other options as well.

To get the statistical properties of a population, we performed a linear interpolation on the final physical quantities used in the SN prescriptions. These quantities are needed to determine the explodability outcome and the H-rich envelope mass  
for a population of 10,000 single stars, drawn from a \citet{Kroupa2001} initial mass function over a range of $8-120 \Msun$.

\section{Results}\label{sec:results}

Figure~\ref{fig:Mfinal} shows the final total, He-core, and CO-core masses as a function of the initial masses of the stars. 
The final total and core  masses are determined by the interplay between (i) the increasing
mass-loss rates, (ii) the shorter lifetime, and (iii) the bigger mass reservoir and the relatively more massive cores of stars with higher initial masses. 
We see  that, on average, stars with higher initial masses have higher
final total and core masses. For our default grid of stellar simulations (Fig.~\ref{fig:Mfinal}a), stars with initial masses of approximately $70\Msun$ reach
a peak final mass of $\sim 35 \Msun$. For even higher initial masses, the final mass drops to around $13\Msun$ due to the increased wind mass-loss rate.

For a fixed set of stellar physics assumptions, there is a  minimum initial-mass threshold, $M_{\rm min,WR}$, above which single stars lose their H-rich envelopes before collapse, making them candidates for exploding as SNe Ib/c.
Type IIb SNe can occur in a thin mass range between $M_{\rm min,IIb}$ and $M_{\rm min,WR}$. These parameters are sensitive to stellar physical assumptions, especially the strength of stellar winds. In our default models, we find $M_{\rm min,IIb} \sim 25 $\Msun and  $M_{\rm min,WR} \sim 30\Msun$, consistent with previous
theoretical and observational estimates
that place $M_{\rm min,WR}$ between $20$ and $40 \Msun$
\citep{Heger+2003,Crowther+2006,Georgy+2012, Renzo+2017, Sravan+2019}. 
In Table \ref{table:results} we report on the values of these parameters and the formation rates of \SNI and \SNIIb candidates for all our stellar simulations.

In Fig.~\ref{fig:Mfinal} we also depict the range of initial masses that lead to successful SNe for various SN engines (horizontal, colored bands). FR12+rapid and FR12+delayed engines (magenta and brown, respectively), which are solely based on the final mass of the carbon-oxygen core (blue line), result in explosions only for initial stellar masses below $\sim 28 \Msun$ in our default model.  Only a few SN IIb  candidates are in this range and successfully explode, whereas all fully stripped candidates of SNe Ib/c are predicted to collapse into a BH. 

S16+N20 and S16+19.8 (gray and light green, respectively) 
as well as P20E16+N20 (orange) show a more abrupt behavior, as expected from the nonlinearity of the inner core structure with initial mass \citep{Sukhbold+2016, Ertl+2016}. 
These SN engines disfavor explosions of initially very massive stars and do not predict any SNe Ib/c or even SNe IIb.

In Table~\ref{table:results} we also show the expected fraction of candidates of \SE that will  successfully explode , following a \citet{Kroupa2001} initial mass function,  
for all the combinations of different physical assumptions during the
stellar evolution (columns) and SN engines (rows). 
None of the explosion models, expect that from C20, produce any SNe Ib/c, regardless of the
  assumptions during the stellar evolution, with the exception of the increased
  wind mass-loss rate as we discuss below. 
F12+rapid and F12+delayed in our default model predict that $\sim15\%$ of partially stripped candidates (with $0 < M_{\rm env} \leq 0.5 \Msun$) will successfully explode as SNe IIb. This fraction is sensitive to the assumed maximum $M_{\rm env}$, increasing to $\gtrsim 60\%$ for F12+rapid and F12+delayed when the assumed range for SNe IIb is  $0 \leq M_{\rm env} < 1.0 \Msun$, although the fraction of successful SNe IIb remains negligible for the other engines, even in this case.

The C20 engine predicts a very different picture compared with the other SN prescriptions, resulting in all SESN progenitors exploding. This is because in that study they find that in the models with higher initial stellar masses  of $\gtrsim 30-50$ \Msun from \citet{Sukhbold+2016}, it is easier to make them explode.   These stars have the highest core masses in \citet{Sukhbold+2016},  $\sim 15$ \Msun. Combining this with the higher core masses found in our models leads to explosions from all SESNe for $\alpha_{\rm \Lambda} \geq 1.2$. 
\citet{Couch+2020} find an island of BH formation from initially lower mass progenitors, of around $10-15$ \Msun, which form  $3-4$ \Msun final helium cores; this can be found in our results as well. Their results are drastically different compared to \citet{Sukhbold+2016} and \citet{Ertl+2016}, possibly pointing toward the importance of  turbulence. In a way, turbulence may be the key factor for solving the issue, although it creates artifacts in the SN II and compact object formation landscapes.  
We stress, however, that the C20 model leads to low mass BHs of $\lesssim 10 \Msun$, especially for high $\alpha_{\rm \Lambda}$ values, which is not consistent with the BH masses in some Galactic BH X-ray binaries \citep[e.g.,][]{Miller-Jones+2021}.

The main reason for the dearth of successful SESN explosions for all SN engines apart from that of C20 is the high final core masses.  The exact final values are dependent on stellar parameters, including  stellar winds, convective core overshooting, the inclusion of the effect of shell convective over- and undershooting, and rotation, but these high final masses are generally consistent with other theoretical and observational results \citep[e.g.,][]{Georgy+2012, Spera+2015, Sander+2019}.
The nonoccurrence of SNe Ib/c is a persistent result in our simulations that do not include effects of turbulence in the collapse process, as in C20, independent of mild
initial rotation or the amount of shell overshooting during the post-main-sequence evolution. 

Only in our grid of increased wind strength ($1.5$ times higher compared to our default model; Fig.~\ref{fig:Mfinal}b, last column of Table~\ref{table:results}) do we find a small fraction of fully stripped progenitors that successfully explode, even for engines different than that of C20. Higher wind mass-loss rates result in a decreased $M_{\rm min, WR}$, increasing the number of SESN candidates that originate from lower initial masses compared to the default simulations. It also leads to less massive final cores compared to default simulations, which on average favors successful explosions and neutron star formations according to all SN engines.  We additionally find successful SNe Ib/c originating from stars of $\gtrsim 60\Msun$ with final masses of $\sim 7-8 \Msun$, although this group is very sensitive to wind mass-loss uncertainties. A higher fraction of SN IIb candidates also explode for models with increased winds. Still, even in this case, the fraction of successful \SNI explosions compared to all fully stripped cadidate stars remains below 5\% for SN engines that take the compactness of the inner core into account (S16+N20, S16+19.8, and Pat20E16+N20). It is interesting to note that the C20 engine in the case of high winds does not lead to successful explosions from all stripped-envelope candidates, which was the case with all other stellar assumptions. 
This is because the lower final cores of these stellar models are found in the mass range in which inducing explosions is found to be difficult when the turbulence effect is included.

 \begin{table*}
\caption{Minimum initial mass for SN Ib/c and IIb candidate stars ($M_{\rm min,WR}$ and $M_{\rm min,IIb}$, respectively), their formation rate for each  stellar simulation, and the fraction of them that  successfully explode in  SNe Ib/c and $\rm{IIb}$, respectively, for different SN engines.
\label{table:results}}.
    \centering
    \begin{tabular}{c||c|c|c|c|c}

        & \multicolumn{4}{c}{Stellar simulation}\\
        \\
           &  default  &  Shell over/& Lower core & Initial rotation, &  Increased winds,  \\
        & (Fig.~\ref{fig:Mfinal}a)  &  undershooting & overshooting & $\omega/\omega_{\rm crit} = 0.4$  &  $\eta=1.5$ (Fig.~\ref{fig:Mfinal}b)\\
        \hline
        $M_{\rm min,WR} (M_{\rm min,IIb}) \, [\Msun]$ &  31 (26.2) & 30 (25.4)  & 55 (53.7) & 27.0 (24.3) & 23 (18.8)  \\
     Ib/c (IIb) cand. $[10^{-3}(\Msun/\rm{yr})^{-1}]$ &  1.3 (0.47) & 1.47 (0.27)  & 0.43 (0.11) &  1.38 (0.61) & 2.5 (0.85)  \\
        \hline
        FR12+rapid & 0\%\,(31\%)& 0\%\,(59\%)& 0\%\,(13.4\%) & 0\%\,(53\%)& 48\%\,(42\%) \\
        FR12+delayed & 0\%\,(30\%) & 0\%\,(68\%)& 0\%\,(7\%) & 0\%\,(52\%)& 65\%\,(100\%) \\
        S16+N20 &  0\%\,(0\%) & 0\%\,(0\%)&  0\%\,(7\%) & 0\%\,(0\%)& 2.6\%\,(44\%)  \\
        S16+S19.8 &  0\%\,(0\%) & 0\%\,(0\%)& 0\%\,(7\%)& 0\%\,(0\%) & 4\%\,(50\%) \\
        P20E16+N20 &  0\%\,(0\%) & 0\%\,(0\%)& 0\%\,(7\%) & 0\%\,(0\%)& 1\%\,(21\%) \\
        C20 & 100\%\,(100\%) &  100\%\,(100\%) & 100\%\,(100\%) & 100\%\,(87\%)& 60\%\,(74\%) \\
    \end{tabular}
\end{table*}

\section{Discussion and conclusions}
\label{sec:conclusions}

 In this letter we estimate the effect of the non-explodability of massive single-star SN
 progenitors on the statistics of SESNe, using various up-to-date core-collapse prescriptions.
 We find that, unless high wind mass loss is invoked, all fully stripped stars that could produce SNe Ib/c will collapse into a BH with a faint transient, or none at all, for all but one of the SN engines considered in this study. A significant fraction of single-star candidate progenitors of IIb SNe are expected to not produce
transient events either. In contrast, the fraction of
stripped-envelope candidates that successfully explode is as high as
100\% for the C20 SN engine that includes the effects of turbulence.
%

There are several caveats to our results.
Since our stellar models stop at core carbon depletion, we cannot directly quantify the explodability of each model at the onset of core collapse. We aim to
explore the general trend for a range of masses with various SN engines rather than predict self-consistently
the outcome of the collapse. The mapping of our stellar structure properties to
the outcome of their collapse is less uncertain in the case of
the Pat20E16+N20 engine because this model additionally takes the core carbon abundance into account.
For this mapping, the
main source of uncertainty seems to originate from stellar evolution
processes prior to carbon ignition:
prior mass loss \citep{Renzo+2017}, internal mixing
\citep{Limongi+2018}, and the efficiency of the nuclear reaction rates
that affect the abundances in the carbon core \citep{Tur+2007, Farmer+2020}. We note that this engine predicts the lowest fraction of successful SESN explosions for all stellar simulations.  In
the other SN engines considered, we implicitly assume that the
explodability outcomes from \cite{Fryer+2012} and \citet{Sukhbold+2016} apply for
stars with the same carbon and helium core masses, respectively, despite
their progenitor models being computed using different stellar 
codes and physical assumptions.  

Previous studies have shown that the suggested explodability metrics depend on metallicity and rotation \citep[e.g.,][]{OConnor+2011},
the adopted stellar wind mass-loss rates \citep[even if these only act early during a star's evolution;][]{Renzo+2017}, overshooting
\citep{Davis+2019}, and the occurrence of convective shell mergers \citep[e.g.,][]{Vartanyan+2021}. Other
parameters might play a significant role in shaping the density
structure of the core at the onset of core collapse. Studying the
complex impact of each stellar physics assumption on the core-collapse
outcome is beyond the scope of this letter.

A major caveat common to explodability studies that span a large
parameter range is that these are only feasible with semi-analytic
\citep[e.g.,][]{Fryer+2012, Mandel+2020,Schneider+2021} and/or
1D parametrized explosion simulations \citep[e.g.,][]{OConnor+2011,
  Ertl+2016, Sukhbold+2016, Ebinger+2019, Ertl+2020}. However, it has become
increasingly clear that multidimensional effects (e.g., neutrino-driven convection, standing accretion shock instability, and lepton
emission asymmetry) are key to the success of explosions
\citep[e.g.,][for a review]{Janka2012} and that asymmetries are necessary
to produce natal kicks \citep[e.g.,][]{Janka2013, Janka+2017}. Recent
multidimensional simulations of core collapse can result in
successful shock revival even when 1D parametrized
results might suggest complete fallback in a failed explosion
\citep{Chan+2018, Chan+2020, Ott+2018, Kuroda+2018, Burrows+2019,Vartanyan+2019, Powell+2020}. In the case of a successful shock revival, we might expect a visible electromagnetic transient. In any case, these simulations are
computationally challenging and
expensive, 
and thus there are not yet any prescriptions for an extended mass range of
progenitors. 

In addition, we focused only on the paradigm of neutrino-driven
explosions. While other explosion mechanisms have been proposed
\citep[e.g.,][]{MacFadyen+Woosley1999,Janka2012, Gilkis+2016,Soker2019}, no criterion to assess the outcome
across a large range of initial masses exists to our knowledge. 
We also note that  possible  faint electromagnetic transients in
the case of a ``failed SN'' and BH formation due to the loss of gravitational mass to neutrinos \citep{Fernandez+2018, ivanov:21} seem quite improbable for compact, stripped-envelope stars, for which little to no
ejection is expected. 

In this study we have restricted ourselves
to 
solar metallicity models; as such, our results do not apply directly to the
observed SESN sample, as these include events from a range of metallicity
environments 
\citep[e.g.,][]{Graur+2017}.  However, in lower metallicities, SN IIb
and Ib/c stellar candidates would be on average less numerous and of higher
final mass due to their lower  wind mass-loss rates \citep[e.g.,][]{Vink+2001,Smith2014}. Therefore, these progenitors would also have on
  average larger final core masses that correspond to less
  explosive structures for most of the SN engines discussed here.

We show that one possible way to increase the number of SESN progenitors from single
stars at a given metallicity would be to increase the stellar wind mass-loss rates 
\citep[e.g.,][]{van-Loon2006,Beasor+Davies2018}.
However, observational evidence and
theoretical arguments thus far point toward potentially lower wind
mass-loss rates than adopted in this study \citep[e.g.,][]{Smith2014, Neijssel+2021, Higgins+2021}. In addition, changing the wind mass-loss rates can
have an impact on the core structure and its explodability
\citep[e.g.,][]{Renzo+2017}. %
Another
possibility is for eruptive mass loss \citep[luminous blue variable eruptions; e.g.,][]{Conti+1977} or episodic mass loss linked to the
late phases of massive stellar evolution \citep[e.g.,][]{Arnett+Meakin2011, Quataert+Shiode2012} to strip a star of its final H-rich envelope, although these processes may instead lead to Type IIn SNe due to interaction with circumstellar material.


If indeed only a small portion of massive, single, stripped stars
contribute to the statistics of SESNe, it would significantly change
their theoretically expected rates and properties. As we cannot empirically constrain the rate of failed SNe from WR stars, their explodability fraction is not an observationally testable parameter but will affect the observed rate of SESN transients. Assuming that all
SESNe originate from single stars, our finding -- that the majority of single stars that end up stripped of their H-rich envelopes do not explode -- is inconsistent with their observed relative rate compared to hydrogen-rich Type II
events of  $\sim 1/3$ 
\citep{Smith+2011,Eldridge+2013, Shivvers+2019}. This is because theoretically expected single SESN  progenitors are very rare,
even if we assume that all massive stars are able to explode ($\sim 0.15$ to all SNe, for  $M_{\rm min,WR} = 31$ \Msun as our default model and assuming Type II explosions from stars with initial masses of between 8 and 31 \Msun ). So the findings
discussed in this study aggravate the possible rate discrepancy \citep[something that is qualitatively mentioned in][]{Smith+2011}. 
The empirical absolute rate of SESNe depends on the star-forming rate of the host galaxies of the progenitors and thus is not easily constrained; however, the fact that we do detect these types of transients obviously cannot be explained by single massive single-star progenitors if their explodability rate is $\sim 0\%$.

Our results alone are not necessarily a motivation to revise single-star mass-loss rates or their explodability properties, since the  binary progenitors of SESNe are a suggested alternative path for  SNe Ib/c and SNe IIb \citep[e.g.,][]{Podsiadlowski+1992, De-Donder+1998, Yoon+2010,Eldridge+2013,Zapartas+2017b,Sravan+2019, Sravan+2020}.
This scenario helps explain their low ejecta masses and short timescales \citep[e.g.,][]{Drout+2011, Lyman+2016, Modjaz+2016}, 
their relatively high occurrence rate \citep[e.g.,][]{Eldridge+2013,Graur+2017,Shivvers+2019}, the difficulty to directly image their progenitors \citep[e.g.,][although see \citealt{Yoon+2012,van-Dyk+2019}]{Eldridge+2013}, and the detection of binary companions at SN IIb sites  \citep{Maund+2004,Fox+2014,Ryder+2018}. 
Within the framework of this study,
our results suggest
that binary progenitors may be the dominant formation channel for \SE. Evolution in a binary system offers two advantages over single-star
progenitors. First, binary interactions do not have a minimum mass
threshold for removing the H-rich envelope \citep{Podsiadlowski+1992, Eldridge+2013, Yoon+2017, Gotberg+2018, Sravan+2018, Sravan+2020}, allowing progenitors of initial masses of $\sim10-25$ \Msun (which have lower final cores and thus will explode in most cases) to produce \SE. Secondly, binary
stripping seems to affect the final core structure and chemical composition, on average increasing the explodability of binary stripped stars compared to single stars of
the same initial mass or even the same carbon core mass
\citep{Schneider+2021, Laplace+2021, Vartanyan+2021}.  
Binaries might also provide paths to \SE through mergers or by increasing the mass of binary secondaries enough for them to wind-strip \citep[e.g.,][]{Yoon+2012,Zapartas+2017b,Hirai+2020}. However, the explodability of such progenitors requires further study.

\begin{acknowledgements}
We thank Sean Couch for sharing the data to reproduce  \citet{Couch+2020} SN prescription.
EZ acknowledges support from 
the Swiss Government Excellence Scholarship (ESKAS No. 2019.0091). This work was supported by the Swiss National Science Foundation Professorship grant (project number PP00P2 176868; PI Tassos Fragos).
JJA and SC are supported by CIERA and AD, JGS, and KAR are supported by the Gordon and Betty Moore Foundation through grant GBMF8477. The computations were performed in part at the University of Geneva on the Baobab and Yggdrasil computer clusters and at Northwestern University on the Trident computer cluster (the latter funded by grant GBMF8477).

\end{acknowledgements}

\bibliographystyle{aa}
\bibliography{my_bib,bib_overleaf}

\begin{thebibliography}{116}
\expandafter\ifx\csname natexlab\endcsname\relax\def\natexlab#1{#1}\fi

\bibitem[{{Abbott} {et~al.}(2021){Abbott}, {Abbott}, {Abraham}, {Acernese},
  {Ackley}, {Adams}, {Adams}, {Adhikari}, {Adya}, {Affeldt}, {Agathos},
  {Agatsuma}, {Aggarwal}, {Aguiar}, {Aiello}, {Ain}, {Ajith}, {Allen},
  {Allocca}, {Altin}, {Amato}, {Anand}, {Ananyeva}, {Anderson}, {Anderson},
  {Angelova}, {Ansoldi}, {Antelis}, {Antier}, {Appert}, {Arai}, {Araya},
  {Areeda}, {Ar{\`e}ne}, {Arnaud}, {Aronson}, {Arun}, {Asali}, {Ascenzi},
  {Ashton}, {Aston}, {Astone}, {Aubin}, {Aufmuth}, {AultONeal}, {Austin},
  {Avendano}, {Babak}, {Badaracco}, {Bader}, {Bae}, {Baer}, {Bagnasco},
  {Baird}, {Ball}, {Ballardin}, {Ballmer}, {Bals}, {Balsamo}, {Baltus},
  {Banagiri}, {Bankar}, {Bankar}, {Barayoga}, {Barbieri}, {Barish}, {Barker},
  {Barneo}, {Barnum}, {Barone}, {Barr}, {Barsotti}, {Barsuglia}, {Barta},
  {Bartlett}, {Bartos}, {Bassiri}, {Basti}, {Bawaj}, {Bayley}, {Bazzan},
  {Becher}, {B{\'e}csy}, {Bedakihale}, {Bejger}, {Belahcene}, {Beniwal},
  {Benjamin}, {Bennett}, {Bentley}, {Bergamin}, {Berger}, {Bergmann},
  {Bernuzzi}, {Berry}, {Bersanetti}, {Bertolini}, {Betzwieser}, {Bhandare},
  {Bhandari}, {Bhattacharjee}, {Bidler}, {Bilenko}, {Billingsley}, {Birney},
  {Birnholtz}, {Biscans}, {Bischi}, {Biscoveanu}, {Bisht}, {Bitossi},
  {Bizouard}, {Blackburn}, {Blackman}, {Blair}, {Blair}, {Blair}, {Blanch},
  {Bobba}, {Bode}, {Boer}, {Boetzel}, {Bogaert}, {Boldrini}, {Bondu},
  {Bonilla}, {Bonnand}, {Booker}, {Boom}, {Bork}, {Boschi}, {Bose},
  {Bossilkov}, {Boudart}, {Bouffanais}, {Bozzi}, {Bradaschia}, {Brady},
  {Bramley}, {Branchesi}, {Brau}, {Breschi}, {Briant}, {Briggs}, {Brighenti},
  {Brillet}, {Brinkmann}, {Brockill}, {Brooks}, {Brooks}, {Brown}, {Brunett},
  {Bruno}, {Bruntz}, {Buikema}, {Bulik}, {Bulten}, {Buonanno}, {Buscicchio},
  {Buskulic}, {Byer}, {Cabero}, {Cadonati}, {Caesar}, {Cagnoli}, {Cahillane},
  {Calder{\'o}n Bustillo}, {Callaghan}, {Callister}, {Calloni}, {Camp},
  {Canepa}, {Cannon}, {Cao}, {Cao}, {Carapella}, {Carbognani}, {Carney},
  {Carpinelli}, {Carullo}, {Carver}, {Casanueva Diaz}, {Casentini}, {Caudill},
  {Cavagli{\`a}}, {Cavalier}, {Cavalieri}, {Cella}, {Cerd{\'a}-Dur{\'a}n},
  {Cesarini}, {Chaibi}, {Chakravarti}, {Chan}, {Chan}, {Chandra}, {Chanial},
  {Chao}, {Charlton}, {Chase}, {Chassande-Mottin}, {Chatterjee},
  {Chattopadhyay}, {Chaturvedi}, {Chatziioannou}, {Chen}, {Chen}, {Chen},
  {Chen}, {Cheng}, {Cheong}, {Chia}, {Chiadini}, {Chierici}, {Chincarini},
  {Chiummo}, {Cho}, {Cho}, {Cho}, {Choate}, {Christensen}, {Chu}, {Chua},
  {Chung}, {Chung}, {Ciani}, {Ciecielag}, {Cie{\'s}lar}, {Cifaldi}, {Ciobanu},
  {Ciolfi}, {Cipriano}, {Cirone}, {Clara}, {Clark}, {Clark}, {Clarke},
  {Clearwater}, {Clesse}, {Cleva}, {Coccia}, {Cohadon}, {Cohen}, {Colleoni},
  {Collette}, {Collins}, {Colpi}, {Constancio}, {Conti}, {Cooper}, {Corban},
  {Corbitt}, {Cordero-Carri{\'o}n}, {Corezzi}, {Corley}, {Cornish}, {Corre},
  {Corsi}, {Cortese}, {Costa}, {Cotesta}, {Coughlin}, {Coughlin}, {Coulon},
  {Countryman}, {Couvares}, {Covas}, {Coward}, {Cowart}, {Coyne}, {Coyne},
  {Creighton}, {Creighton}, {Croquette}, {Crowder}, {Cudell}, {Cullen},
  {Cumming}, {Cummings}, {Cunningham}, {Cuoco}, {Curylo}, {Dal Canton},
  {D{\'a}lya}, {Dana}, {DaneshgaranBajastani}, {D'Angelo}, {Danilishin},
  {D'Antonio}, {Danzmann}, {Darsow-Fromm}, {Dasgupta}, {Datrier}, {Dattilo},
  {Dave}, {Davier}, {Davies}, {Davis}, {Daw}, {Dean}, {DeBra}, {Deenadayalan},
  {Degallaix}, {De Laurentis}, {Del{\'e}glise}, {Del Favero}, {De Lillo}, {De
  Lillo}, {Del Pozzo}, {DeMarchi}, {De Matteis}, {D'Emilio}, {Demos}, {Denker},
  {Dent}, {Depasse}, {De Pietri}, {De Rosa}, {De Rossi}, {DeSalvo}, {de
  Varona}, {Dhurandhar}, {D{\'\i}az}, {Diaz-Ortiz}, {Didio}, {Dietrich}, {Di
  Fiore}, {DiFronzo}, {Di Giorgio}, {Di Giovanni}, {Di Giovanni}, {Di
  Girolamo}, {Di Lieto}, {Ding}, {Di Pace}, {Di Palma}, {Di Renzo},
  {Divakarla}, {Dmitriev}, {Doctor}, {D'Onofrio}, {Donovan}, {Dooley},
  {Doravari}, {Dorrington}, {Downes}, {Drago}, {Driggers}, {Du}, {Ducoin},
  {Dupej}, {Durante}, {D'Urso}, {Duverne}, {Dwyer}, {Easter}, {Eddolls},
  {Edelman}, {Edo}, {Edy}, {Effler}, {Eichholz}, {Eikenberry}, {Eisenmann},
  {Eisenstein}, {Ejlli}, {Errico}, {Essick}, {Estell{\'e}s}, {Estevez},
  {Etienne}, {Etzel}, {Evans}, {Evans}, {Ewing}, {Fafone}, {Fair}, {Fairhurst},
  {Fan}, {Farah}, {Farinon}, {Farr}, {Farr}, {Fauchon-Jones}, {Favata}, {Fays},
  {Fazio}, {Feicht}, {Fejer}, {Feng}, {Fenyvesi}, {Ferguson},
  {Fernandez-Galiana}, {Ferrante}, {Ferreira}, {Fidecaro}, {Figura}, {Fiori},
  {Fiorucci}, {Fishbach}, {Fisher}, {Fishner}, {Fittipaldi}, {Fitz-Axen},
  {Fiumara}, {Flaminio}, {Floden}, {Flynn}, {Fong}, {Font}, {Forsyth},
  {Fournier}, {Frasca}, {Frasconi}, {Frei}, {Freise}, {Frey}, {Frey},
  {Fritschel}, {Frolov}, {Fronz{\'e}}, {Fulda}, {Fyffe}, {Gabbard}, {Gadre},
  {Gaebel}, {Gair}, {Gais}, {Galaudage}, {Gamba}, {Ganapathy}, {Ganguly},
  {Gaonkar}, {Garaventa}, {Garc{\'\i}a-Quir{\'o}s}, {Garufi}, {Gateley},
  {Gaudio}, {Gayathri}, {Gemme}, {Gennai}, {George}, {George}, {Gergely},
  {Ghonge}, {Ghosh}, {Ghosh}, {Ghosh}, {Giacomazzo}, {Giacoppo}, {Giaime},
  {Giardina}, {Gibson}, {Gier}, {Gill}, {Giri}, {Glanzer}, {Gleckl}, {Godwin},
  {Goetz}, {Goetz}, {Gohlke}, {Goncharov}, {Gonz{\'a}lez}, {Gopakumar},
  {Gossan}, {Gosselin}, {Gouaty}, {Grace}, {Grado}, {Granata}, {Granata},
  {Grant}, {Gras}, {Grassia}, {Gray}, {Gray}, {Greco}, {Green}, {Green},
  {Gretarsson}, {Griggs}, {Grignani}, {Grimaldi}, {Grimes}, {Grimm}, {Grote},
  {Grunewald}, {Gruning}, {Guerrero}, {Guidi}, {Guimaraes}, {Guix{\'e}},
  {Gulati}, {Guo}, {Gupta}, {Gupta}, {Gupta}, {Gustafson}, {Gustafson},
  {Guzman}, {Haegel}, {Halim}, {Hall}, {Hamilton}, {Hammond}, {Haney}, {Hanke},
  {Hanks}, {Hanna}, {Hannuksela}, {Hannuksela}, {Hansen}, {Hansen}, {Hanson},
  {Harder}, {Hardwick}, {Haris}, {Harms}, {Harry}, {Harry}, {Hartwig},
  {Hasskew}, {Haster}, {Haughian}, {Hayes}, {Healy}, {Heidmann}, {Heintze},
  {Heinze}, {Heinzel}, {Heitmann}, {Hellman}, {Hello}, {Helmling-Cornell},
  {Hemming}, {Hendry}, {Heng}, {Hennes}, {Hennig}, {Hennig}, {Hernandez
  Vivanco}, {Heurs}, {Hild}, {Hill}, {Hines}, {Hochheim}, {Hofgard}, {Hofman},
  {Hohmann}, {Holgado}, {Holland}, {Hollows}, {Holmes}, {Holt}, {Holz},
  {Hopkins}, {Horst}, {Hough}, {Howell}, {Hoy}, {Hoyland}, {Huang},
  {H{\"u}bner}, {Huddart}, {Huerta}, {Hughey}, {Hui}, {Husa}, {Huttner},
  {Hutzler}, {Huxford}, {Huynh-Dinh}, {Idzkowski}, {Iess}, {Imperato},
  {Inchauspe}, {Ingram}, {Intini}, {Isi}, {Iyer}, {JaberianHamedan}, {Jacqmin},
  {Jadhav}, {Jadhav}, {James}, {Jani}, {Janssens}, {Janthalur}, {Jaranowski},
  {Jariwala}, {Jaume}, {Jenkins}, {Jeunon}, {Jiang}, {Johns}, {Jones}, {Jones},
  {Jones}, {Jones}, {Jones}, {Jonker}, {Ju}, {Junker}, {Kalaghatgi},
  {Kalogera}, {Kamai}, {Kandhasamy}, {Kang}, {Kanner}, {Kapadia}, {Kapasi},
  {Karathanasis}, {Karki}, {Kashyap}, {Kasprzack}, {Kastaun}, {Katsanevas},
  {Katsavounidis}, {Katzman}, {Kawabe}, {K{\'e}f{\'e}lian}, {Keitel}, {Key},
  {Khadka}, {Khalili}, {Khan}, {Khan}, {Khazanov}, {Khetan}, {Khursheed},
  {Kijbunchoo}, {Kim}, {Kim}, {Kim}, {Kim}, {Kim}, {Kim}, {Kimball}, {King},
  {Kinley-Hanlon}, {Kirchhoff}, {Kissel}, {Kleybolte}, {Klimenko}, {Knowles},
  {Knyazev}, {Koch}, {Koehlenbeck}, {Koekoek}, {Koley}, {Kolstein}, {Komori},
  {Kondrashov}, {Kontos}, {Koper}, {Korobko}, {Korth}, {Kovalam}, {Kozak},
  {Kr{\"a}mer}, {Kringel}, {Krishnendu}, {Kr{\'o}lak}, {Kuehn}, {Kumar},
  {Kumar}, {Kumar}, {Kumar}, {Kuns}, {Kwang}, {Lackey}, {Laghi}, {Lalande},
  {Lam}, {Lamberts}, {Landry}, {Lane}, {Lang}, {Lange}, {Lantz}, {Lanza}, {La
  Rosa}, {Lartaux-Vollard}, {Lasky}, {Laxen}, {Lazzarini}, {Lazzaro}, {Leaci},
  {Leavey}, {Lecoeuche}, {Lee}, {Lee}, {Lee}, {Lee}, {Lehmann}, {Leon},
  {Leroy}, {Letendre}, {Levin}, {Li}, {Li}, {Li}, {Li}, {Li}, {Linde},
  {Linker}, {Linley}, {Littenberg}, {Liu}, {Liu}, {Llorens-Monteagudo}, {Lo},
  {Lockwood}, {London}, {Longo}, {Lorenzini}, {Loriette}, {Lormand}, {Losurdo},
  {Lough}, {Lousto}, {Lovelace}, {L{\"u}ck}, {Lumaca}, {Lundgren}, {Ma},
  {Macas}, {MacInnis}, {Macleod}, {MacMillan}, {Macquet}, {Maga{\~n}a
  Hernandez}, {Maga{\~n}a-Sandoval}, {Magazz{\`u}}, {Magee}, {Majorana},
  {Maksimovic}, {Maliakal}, {Malik}, {Man}, {Mandic}, {Mangano}, {Mansell},
  {Manske}, {Mantovani}, {Mapelli}, {Marchesoni}, {Marion}, {M{\'a}rka},
  {M{\'a}rka}, {Markakis}, {Markosyan}, {Markowitz}, {Maros}, {Marquina},
  {Marsat}, {Martelli}, {Martin}, {Martin}, {Martinez}, {Martinez}, {Martynov},
  {Masalehdan}, {Mason}, {Massera}, {Masserot}, {Massinger}, {Masso-Reid},
  {Mastrogiovanni}, {Matas}, {Mateu-Lucena}, {Matichard}, {Matiushechkina},
  {Mavalvala}, {Maynard}, {McCann}, {McCarthy}, {McClelland}, {McCormick},
  {McCuller}, {McGuire}, {McIsaac}, {McIver}, {McManus}, {McRae}, {McWilliams},
  {Meacher}, {Meadors}, {Mehmet}, {Mehta}, {Melatos}, {Melchor}, {Mendell},
  {Menendez-Vazquez}, {Mercer}, {Mereni}, {Merfeld}, {Merilh}, {Merritt},
  {Merzougui}, {Meshkov}, {Messenger}, {Messick}, {Metzdorff}, {Meyers},
  {Meylahn}, {Mhaske}, {Miani}, {Miao}, {Michaloliakos}, {Michel}, {Middleton},
  {Milano}, {Miller}, {Miller}, {Millhouse}, {Mills}, {Milotti},
  {Milovich-Goff}, {Minazzoli}, {Minenkov}, {Mir}, {Mishkin}, {Mishra},
  {Mistry}, {Mitra}, {Mitrofanov}, {Mitselmakher}, {Mittleman}, {Mo},
  {Mogushi}, {Mohapatra}, {Mohite}, {Molina}, {Molina-Ruiz}, {Mondin},
  {Montani}, {Moore}, {Moraru}, {Morawski}, {Moreno}, {Morisaki}, {Mours},
  {Mow-Lowry}, {Mozzon}, {Muciaccia}, {Mukherjee}, {Mukherjee}, {Mukherjee},
  {Mukherjee}, {Mukund}, {Mullavey}, {Munch}, {Mu{\~n}iz}, {Murray}, {Nadji},
  {Nagar}, {Nardecchia}, {Naticchioni}, {Nayak}, {Neil}, {Neilson}, {Nelemans},
  {Nelson}, {Nery}, {Neunzert}, {Ng}, {Ng}, {Nguyen}, {Nguyen}, {Nguyen},
  {Nichols}, {Nissanke}, {Nocera}, {Noh}, {North}, {Nothard}, {Nuttall},
  {Oberling}, {O'Brien}, {O'Dell}, {Oganesyan}, {Ogin}, {Oh}, {Oh}, {Ohme},
  {Ohta}, {Okada}, {Olivetto}, {Oppermann}, {Oram}, {O'Reilly}, {Ormiston},
  {Ormsby}, {Ortega}, {O'Shaughnessy}, {Ossokine}, {Osthelder}, {Ottaway},
  {Overmier}, {Owen}, {Pace}, {Pagano}, {Page}, {Pagliaroli}, {Pai}, {Pai},
  {Palamos}, {Palashov}, {Palomba}, {Pan}, {Panda}, {Pang}, {Pankow},
  {Pannarale}, {Pant}, {Paoletti}, {Paoli}, {Paolone}, {Parker}, {Pascucci},
  {Pasqualetti}, {Passaquieti}, {Passuello}, {Patel}, {Patricelli}, {Payne},
  {Pechsiri}, {Pedraza}, {Pegoraro}, {Pele}, {Penn}, {Perego}, {Perez},
  {P{\'e}rigois}, {Perreca}, {Perri{\`e}s}, {Petermann}, {Petterson},
  {Pfeiffer}, {Pham}, {Phukon}, {Piccinni}, {Pichot}, {Piendibene},
  {Piergiovanni}, {Pierini}, {Pierro}, {Pillant}, {Pilo}, {Pinard}, {Pinto},
  {Piotrzkowski}, {Pirello}, {Pitkin}, {Placidi}, {Plastino}, {Pluchar},
  {Poggiani}, {Polini}, {Pong}, {Ponrathnam}, {Popolizio}, {Porter},
  {Poverman}, {Powell}, {Pracchia}, {Prajapati}, {Prasai}, {Prasanna},
  {Pratten}, {Prestegard}, {Principe}, {Prodi}, {Prokhorov}, {Prosposito},
  {Puecher}, {Punturo}, {Puosi}, {Puppo}, {P{\"u}rrer}, {Qi}, {Quetschke},
  {Quinonez}, {Quitzow-James}, {Raab}, {Raaijmakers}, {Radkins}, {Radulesco},
  {Raffai}, {Rafferty}, {Rail}, {Raja}, {Rajan}, {Rajbhandari}, {Rakhmanov},
  {Ramirez}, {Ramirez}, {Ramos-Buades}, {Rana}, {Rao}, {Rapagnani}, {Rapol},
  {Ratto}, {Raymond}, {Razzano}, {Read}, {Regimbau}, {Rei}, {Reid}, {Reitze},
  {Rettegno}, {Ricci}, {Richardson}, {Richardson}, {Richardson}, {Ricker},
  {Riemenschneider}, {Riles}, {Rizzo}, {Robertson}, {Robinet}, {Rocchi},
  {Rocha}, {Rodriguez}, {Rodriguez-Soto}, {Rolland}, {Rollins}, {Roma},
  {Romanelli}, {Romano}, {Romel}, {Romero}, {Romero-Shaw}, {Romie}, {Ronchini},
  {Rose}, {Rose}, {Rose}, {Rosell}, {Rosi{\'n}ska}, {Rosofsky}, {Ross},
  {Rowan}, {Rowlinson}, {Roy}, {Roy}, {Ruggi}, {Ryan}, {Sachdev}, {Sadecki},
  {Sakellariadou}, {Salafia}, {Salconi}, {Saleem}, {Samajdar}, {Sanchez},
  {Sanchez}, {Sanchez}, {Sanchis-Gual}, {Sanders}, {Santiago}, {Santos},
  {Saravanan}, {Sarin}, {Sassolas}, {Sathyaprakash}, {Sauter}, {Savage},
  {Savant}, {Sawant}, {Sayah}, {Schaetzl}, {Schale}, {Scheel}, {Scheuer},
  {Schindler-Tyka}, {Schmidt}, {Schnabel}, {Schofield}, {Sch{\"o}nbeck},
  {Schreiber}, {Schulte}, {Schutz}, {Schwarm}, {Schwartz}, {Scott}, {Scott},
  {Seglar-Arroyo}, {Seidel}, {Sellers}, {Sengupta}, {Sennett}, {Sentenac},
  {Sequino}, {Sergeev}, {Setyawati}, {Shaffer}, {Shahriar}, {Sharifi},
  {Sharma}, {Sharma}, {Shawhan}, {Shen}, {Shikauchi}, {Shink}, {Shoemaker},
  {Shoemaker}, {Shukla}, {ShyamSundar}, {Sieniawska}, {Sigg}, {Singer},
  {Singh}, {Singh}, {Singha}, {Singhal}, {Sintes}, {Sipala}, {Skliris},
  {Slagmolen}, {Slaven-Blair}, {Smetana}, {Smith}, {Smith}, {Somala}, {Son},
  {Soni}, {Sorazu}, {Sordini}, {Sorrentino}, {Sorrentino}, {Soulard},
  {Souradeep}, {Sowell}, {Spencer}, {Spera}, {Srivastava}, {Srivastava},
  {Staats}, {Stachie}, {Steer}, {Steinke}, {Steinlechner}, {Steinlechner},
  {Steinmeyer}, {Stevenson}, {Stolle-McAllister}, {Stops}, {Stover}, {Strain},
  {Stratta}, {Strunk}, {Sturani}, {Stuver}, {S{\"u}dbeck}, {Sudhagar},
  {Sudhir}, {Suh}, {Summerscales}, {Sun}, {Sun}, {Sunil}, {Sur}, {Suresh},
  {Sutton}, {Swinkels}, {Szczepa{\'n}czyk}, {Tacca}, {Tait}, {Talbot},
  {Tanasijczuk}, {Tanner}, {Tao}, {Tapia}, {Tapia San Martin}, {Tasson},
  {Taylor}, {Tenorio}, {Terkowski}, {Thirugnanasambandam}, {Thomas}, {Thomas},
  {Thomas}, {Thompson}, {Thondapu}, {Thorne}, {Thrane}, {Tiwari}, {Tiwari},
  {Tiwari}, {Toland}, {Tolley}, {Tonelli}, {Tornasi}, {Torres-Forn{\'e}},
  {Torrie}, {Tosta e Melo}, {T{\"o}yr{\"a}}, {Tran}, {Trapananti}, {Travasso},
  {Traylor}, {Tringali}, {Tripathee}, {Trovato}, {Trudeau}, {Tsai}, {Tsang},
  {Tse}, {Tso}, {Tsukada}, {Tsuna}, {Tsutsui}, {Turconi}, {Ubhi}, {Udall},
  {Ueno}, {Ugolini}, {Unnikrishnan}, {Urban}, {Usman}, {Utina}, {Vahlbruch},
  {Vajente}, {Vajpeyi}, {Valdes}, {Valentini}, {Valsan}, {van Bakel},
  {Beuzekom}, {van den Brand}, {Van Den Broeck}, {Vander-Hyde}, {van der
  Schaaf}, {van Heijningen}, {Vardaro}, {Vargas}, {Varma}, {Vass},
  {Vas{\'u}th}, {Vecchio}, {Vedovato}, {Veitch}, {Veitch}, {Venkateswara},
  {Venneberg}, {Venugopalan}, {Verkindt}, {Verma}, {Veske}, {Vetrano},
  {Vicer{\'e}}, {Viets}, {Villa-Ortega}, {Vinet}, {Vitale}, {Vo}, {Vocca},
  {Vorvick}, {Vyatchanin}, {Wade}, {Wade}, {Wade}, {Walet}, {Walker},
  {Wallace}, {Wallace}, {Walsh}, {Wang}, {Wang}, {Wang}, {Wang}, {Ward},
  {Warner}, {Was}, {Washington}, {Watchi}, {Weaver}, {Wei}, {Weinert},
  {Weinstein}, {Weiss}, {Wellmann}, {Wen}, {We{\ss}els}, {Westhouse}, {Wette},
  {Whelan}, {White}, {White}, {Whiting}, {Whittle}, {Wilken}, {Williams},
  {Williams}, {Williamson}, {Willis}, {Willke}, {Wilson}, {Wimmer}, {Winkler},
  {Wipf}, {Woan}, {Woehler}, {Wofford}, {Wong}, {Wrangel}, {Wright}, {Wu},
  {Wysocki}, {Xiao}, {Yamamoto}, {Yang}, {Yang}, {Yang}, {Yap}, {Yeeles},
  {Yoon}, {Yu}, {Yu}, {Yuen}, {Zadro{\.z}ny}, {Zanolin}, {Zelenova}, {Zendri},
  {Zevin}, {Zhang}, {Zhang}, {Zhang}, {Zhang}, {Zhao}, {Zhao}, {Zhou}, {Zhou},
  {Zhu}, {Zimmerman}, {Zucker}, {Zweizig}, {LIGO Scientific Collaboration}, \&
  {Virgo Collaboration}}]{Abbott+2021}
{Abbott}, R., {Abbott}, T.~D., {Abraham}, S., {et~al.} 2021, \apjl, 913, L7

\bibitem[{{Adams} {et~al.}(2017{\natexlab{a}}){Adams}, {Kochanek}, {Gerke}, \&
  {Stanek}}]{Adams+2017b}
{Adams}, S.~M., {Kochanek}, C.~S., {Gerke}, J.~R., \& {Stanek}, K.~Z.
  2017{\natexlab{a}}, \mnras, 469, 1445

\bibitem[{{Adams} {et~al.}(2017{\natexlab{b}}){Adams}, {Kochanek}, {Gerke},
  {Stanek}, \& {Dai}}]{Adams+2017a}
{Adams}, S.~M., {Kochanek}, C.~S., {Gerke}, J.~R., {Stanek}, K.~Z., \& {Dai},
  X. 2017{\natexlab{b}}, \mnras, 468, 4968

\bibitem[{{Antoni} \& {Quataert}(2021)}]{antoni:21}
{Antoni}, A. \& {Quataert}, E. 2021, arXiv e-prints, arXiv:2107.09068

\bibitem[{{Arnett} \& {Meakin}(2011)}]{Arnett+Meakin2011}
{Arnett}, W.~D. \& {Meakin}, C. 2011, \apj, 741, 33

\bibitem[{{Asplund} {et~al.}(2009){Asplund}, {Grevesse}, {Sauval}, \&
  {Scott}}]{Asplund+2009}
{Asplund}, M., {Grevesse}, N., {Sauval}, A.~J., \& {Scott}, P. 2009, \araa, 47,
  481

\bibitem[{{Atri} {et~al.}(2019){Atri}, {Miller-Jones}, {Bahramian}, {Plotkin},
  {Jonker}, {Nelemans}, {Maccarone}, {Sivakoff}, {Deller}, {Chaty}, {Torres},
  {Horiuchi}, {McCallum}, {Natusch}, {Phillips}, {Stevens}, \&
  {Weston}}]{Atri+2019}
{Atri}, P., {Miller-Jones}, J.~C.~A., {Bahramian}, A., {et~al.} 2019, \mnras,
  489, 3116

\bibitem[{{Bavera} {et~al.}(2020){Bavera}, {Fragos}, {Qin}, {Zapartas},
  {Neijssel}, {Mandel}, {Batta}, {Gaebel}, {Kimball}, \&
  {Stevenson}}]{Bavera+2020}
{Bavera}, S.~S., {Fragos}, T., {Qin}, Y., {et~al.} 2020, \aap, 635, A97

\bibitem[{{Beasor} \& {Davies}(2018)}]{Beasor+Davies2018}
{Beasor}, E.~R. \& {Davies}, B. 2018, \mnras, 475, 55

\bibitem[{{Belczynski} {et~al.}(2016){Belczynski}, {Holz}, {Bulik}, \&
  {O'Shaughnessy}}]{Belczynski+2016}
{Belczynski}, K., {Holz}, D.~E., {Bulik}, T., \& {O'Shaughnessy}, R. 2016,
  \nat, 534, 512

\bibitem[{{Bersten} {et~al.}(2012){Bersten}, {Benvenuto}, {Nomoto}, {Ergon},
  {Folatelli}, {Sollerman}, {Benetti}, {Botticella}, {Fraser}, {Kotak},
  {Maeda}, {Ochner}, \& {Tomasella}}]{Bersten+2012}
{Bersten}, M.~C., {Benvenuto}, O.~G., {Nomoto}, K., {et~al.} 2012, \apj, 757,
  31

\bibitem[{{Brott} {et~al.}(2011){Brott}, {Evans}, {Hunter}, {de Koter},
  {Langer}, {Dufton}, {Cantiello}, {Trundle}, {Lennon}, {de Mink}, {Yoon}, \&
  {Anders}}]{Brott+2011a}
{Brott}, I., {Evans}, C.~J., {Hunter}, I., {et~al.} 2011, \aap, 530, A116

\bibitem[{{Burrows} {et~al.}(1995){Burrows}, {Hayes}, \&
  {Fryxell}}]{Burrows+1995}
{Burrows}, A., {Hayes}, J., \& {Fryxell}, B.~A. 1995, \apj, 450, 830

\bibitem[{{Burrows} {et~al.}(2019){Burrows}, {Radice}, \&
  {Vartanyan}}]{Burrows+2019}
{Burrows}, A., {Radice}, D., \& {Vartanyan}, D. 2019, \mnras, 485, 3153

\bibitem[{{Chaboyer} \& {Zahn}(1992)}]{Chaboyer+1992}
{Chaboyer}, B. \& {Zahn}, J.-P. 1992, \aap, 253, 173

\bibitem[{{Chan} {et~al.}(2020){Chan}, {M{\"u}ller}, \& {Heger}}]{Chan+2020}
{Chan}, C., {M{\"u}ller}, B., \& {Heger}, A. 2020, \mnras, 495, 3751

\bibitem[{{Chan} {et~al.}(2018){Chan}, {M{\"u}ller}, {Heger}, {Pakmor}, \&
  {Springel}}]{Chan+2018}
{Chan}, C., {M{\"u}ller}, B., {Heger}, A., {Pakmor}, R., \& {Springel}, V.
  2018, \apjl, 852, L19

\bibitem[{{Choi} {et~al.}(2016){Choi}, {Dotter}, {Conroy}, {Cantiello},
  {Paxton}, \& {Johnson}}]{Choi+2016}
{Choi}, J., {Dotter}, A., {Conroy}, C., {et~al.} 2016, \apj, 823, 102

\bibitem[{{Claret} \& {Torres}(2017)}]{Claret+Torres2017}
{Claret}, A. \& {Torres}, G. 2017, \apj, 849, 18

\bibitem[{{Conti} \& {Ebbets}(1977)}]{Conti+1977}
{Conti}, P.~S. \& {Ebbets}, D. 1977, \apj, 213, 438

\bibitem[{{Couch} {et~al.}(2020){Couch}, {Warren}, \& {O'Connor}}]{Couch+2020}
{Couch}, S.~M., {Warren}, M.~L., \& {O'Connor}, E.~P. 2020, \apj, 890, 127

\bibitem[{{Crowther}(2007)}]{Crowther2007}
{Crowther}, P.~A. 2007, \araa, 45, 177

\bibitem[{{Crowther} {et~al.}(2006){Crowther}, {Hadfield}, {Clark},
  {Negueruela}, \& {Vacca}}]{Crowther+2006}
{Crowther}, P.~A., {Hadfield}, L.~J., {Clark}, J.~S., {Negueruela}, I., \&
  {Vacca}, W.~D. 2006, \mnras, 372, 1407

\bibitem[{Cyburt {et~al.}(2010)Cyburt, Amthor, Ferguson, Meisel, Smith, Warren,
  Heger, Hoffman, Rauscher, Sakharuk, Schatz, Thielemann, \&
  Wiescher}]{Cyburt+2010}
Cyburt, R.~H., Amthor, A.~M., Ferguson, R., {et~al.} 2010, The Astrophysical
  Journal Supplement Series, 189, 240

\bibitem[{{Davis} {et~al.}(2019){Davis}, {Jones}, \& {Herwig}}]{Davis+2019}
{Davis}, A., {Jones}, S., \& {Herwig}, F. 2019, \mnras, 484, 3921

\bibitem[{{De Donder} \& {Vanbeveren}(1998)}]{De-Donder+1998}
{De Donder}, E. \& {Vanbeveren}, D. 1998, \aap, 333, 557

\bibitem[{{de Jager} {et~al.}(1988){de Jager}, {Nieuwenhuijzen}, \& {van der
  Hucht}}]{de-Jager+1988}
{de Jager}, C., {Nieuwenhuijzen}, H., \& {van der Hucht}, K.~A. 1988, \aaps,
  72, 259

\bibitem[{{Dessart} {et~al.}(2012){Dessart}, {Hillier}, {Li}, \&
  {Woosley}}]{Dessart+2012}
{Dessart}, L., {Hillier}, D.~J., {Li}, C., \& {Woosley}, S. 2012, \mnras, 424,
  2139

\bibitem[{{Drout} {et~al.}(2011){Drout}, {Soderberg}, {Gal-Yam}, {Cenko},
  {Fox}, {Leonard}, {Sand}, {Moon}, {Arcavi}, \& {Green}}]{Drout+2011}
{Drout}, M.~R., {Soderberg}, A.~M., {Gal-Yam}, A., {et~al.} 2011, \apj, 741, 97

\bibitem[{{Ebinger} {et~al.}(2019){Ebinger}, {Curtis}, {Fr{\"o}hlich},
  {Hempel}, {Perego}, {Liebend{\"o}rfer}, \& {Thielemann}}]{Ebinger+2019}
{Ebinger}, K., {Curtis}, S., {Fr{\"o}hlich}, C., {et~al.} 2019, \apj, 870, 1

\bibitem[{{Eldridge} {et~al.}(2013){Eldridge}, {Fraser}, {Smartt}, {Maund}, \&
  {Crockett}}]{Eldridge+2013}
{Eldridge}, J.~J., {Fraser}, M., {Smartt}, S.~J., {Maund}, J.~R., \&
  {Crockett}, R.~M. 2013, \mnras, 436, 774

\bibitem[{{Ertl} {et~al.}(2016){Ertl}, {Janka}, {Woosley}, {Sukhbold}, \&
  {Ugliano}}]{Ertl+2016}
{Ertl}, T., {Janka}, H.-T., {Woosley}, S.~E., {Sukhbold}, T., \& {Ugliano}, M.
  2016, \apj, 818, 124

\bibitem[{{Ertl} {et~al.}(2020){Ertl}, {Woosley}, {Sukhbold}, \&
  {Janka}}]{Ertl+2020}
{Ertl}, T., {Woosley}, S.~E., {Sukhbold}, T., \& {Janka}, H.~T. 2020, \apj,
  890, 51

\bibitem[{{Farmer} {et~al.}(2020){Farmer}, {Renzo}, {de Mink}, {Fishbach}, \&
  {Justham}}]{Farmer+2020}
{Farmer}, R., {Renzo}, M., {de Mink}, S.~E., {Fishbach}, M., \& {Justham}, S.
  2020, \apjl, 902, L36

\bibitem[{{Fern{\'a}ndez} {et~al.}(2018){Fern{\'a}ndez}, {Quataert},
  {Kashiyama}, \& {Coughlin}}]{Fernandez+2018}
{Fern{\'a}ndez}, R., {Quataert}, E., {Kashiyama}, K., \& {Coughlin}, E.~R.
  2018, \mnras, 476, 2366

\bibitem[{{Filippenko}(1997)}]{Filippenko1997}
{Filippenko}, A.~V. 1997, \araa, 35, 309

\bibitem[{{Fox} {et~al.}(2014){Fox}, {Azalee Bostroem}, {Van Dyk},
  {Filippenko}, {Fransson}, {Matheson}, {Cenko}, {Chandra}, {Dwarkadas}, {Li},
  {Parker}, \& {Smith}}]{Fox+2014}
{Fox}, O.~D., {Azalee Bostroem}, K., {Van Dyk}, S.~D., {et~al.} 2014, \apj,
  790, 17

\bibitem[{{Fryer} {et~al.}(2012){Fryer}, {Belczynski}, {Wiktorowicz},
  {Dominik}, {Kalogera}, \& {Holz}}]{Fryer+2012}
{Fryer}, C.~L., {Belczynski}, K., {Wiktorowicz}, G., {et~al.} 2012, \apj, 749,
  91

\bibitem[{{Fryer} \& {Heger}(2000)}]{Fryer+Heger2000}
{Fryer}, C.~L. \& {Heger}, A. 2000, \apj, 541, 1033

\bibitem[{{Gaia Collaboration} {et~al.}(2018){Gaia Collaboration}, {Brown},
  {Vallenari}, {Prusti}, {de Bruijne}, {Babusiaux}, {Bailer-Jones}, {Biermann},
  {Evans}, {Eyer}, {Jansen}, {Jordi}, {Klioner}, {Lammers}, {Lindegren},
  {Luri}, {Mignard}, {Panem}, {Pourbaix}, {Randich}, {Sartoretti}, {Siddiqui},
  {Soubiran}, {van Leeuwen}, {Walton}, {Arenou}, {Bastian}, {Cropper},
  {Drimmel}, {Katz}, {Lattanzi}, {Bakker}, {Cacciari}, {Casta{\~n}eda},
  {Chaoul}, {Cheek}, {De Angeli}, {Fabricius}, {Guerra}, {Holl}, {Masana},
  {Messineo}, {Mowlavi}, {Nienartowicz}, {Panuzzo}, {Portell}, {Riello},
  {Seabroke}, {Tanga}, {Th{\'e}venin}, {Gracia-Abril}, {Comoretto},
  {Garcia-Reinaldos}, {Teyssier}, {Altmann}, {Andrae}, {Audard},
  {Bellas-Velidis}, {Benson}, {Berthier}, {Blomme}, {Burgess}, {Busso},
  {Carry}, {Cellino}, {Clementini}, {Clotet}, {Creevey}, {Davidson}, {De
  Ridder}, {Delchambre}, {Dell'Oro}, {Ducourant},
  {Fern{\'a}ndez-Hern{\'a}ndez}, {Fouesneau}, {Fr{\'e}mat}, {Galluccio},
  {Garc{\'\i}a-Torres}, {Gonz{\'a}lez-N{\'u}{\~n}ez}, {Gonz{\'a}lez-Vidal},
  {Gosset}, {Guy}, {Halbwachs}, {Hambly}, {Harrison}, {Hern{\'a}ndez},
  {Hestroffer}, {Hodgkin}, {Hutton}, {Jasniewicz}, {Jean-Antoine-Piccolo},
  {Jordan}, {Korn}, {Krone-Martins}, {Lanzafame}, {Lebzelter}, {L{\"o}ffler},
  {Manteiga}, {Marrese}, {Mart{\'\i}n-Fleitas}, {Moitinho}, {Mora}, {Muinonen},
  {Osinde}, {Pancino}, {Pauwels}, {Petit}, {Recio-Blanco}, {Richards},
  {Rimoldini}, {Robin}, {Sarro}, {Siopis}, {Smith}, {Sozzetti}, {S{\"u}veges},
  {Torra}, {van Reeven}, {Abbas}, {Abreu Aramburu}, {Accart}, {Aerts},
  {Altavilla}, {{\'A}lvarez}, {Alvarez}, {Alves}, {Anderson}, {Andrei},
  {Anglada Varela}, {Antiche}, {Antoja}, {Arcay}, {Astraatmadja}, {Bach},
  {Baker}, {Balaguer-N{\'u}{\~n}ez}, {Balm}, {Barache}, {Barata}, {Barbato},
  {Barblan}, {Barklem}, {Barrado}, {Barros}, {Barstow}, {Bartholom{\'e}
  Mu{\~n}oz}, {Bassilana}, {Becciani}, {Bellazzini}, {Berihuete}, {Bertone},
  {Bianchi}, {Bienaym{\'e}}, {Blanco-Cuaresma}, {Boch}, {Boeche}, {Bombrun},
  {Borrachero}, {Bossini}, {Bouquillon}, {Bourda}, {Bragaglia}, {Bramante},
  {Breddels}, {Bressan}, {Brouillet}, {Br{\"u}semeister}, {Brugaletta},
  {Bucciarelli}, {Burlacu}, {Busonero}, {Butkevich}, {Buzzi}, {Caffau},
  {Cancelliere}, {Cannizzaro}, {Cantat-Gaudin}, {Carballo}, {Carlucci},
  {Carrasco}, {Casamiquela}, {Castellani}, {Castro-Ginard}, {Charlot},
  {Chemin}, {Chiavassa}, {Cocozza}, {Costigan}, {Cowell}, {Crifo}, {Crosta},
  {Crowley}, {Cuypers}, {Dafonte}, {Damerdji}, {Dapergolas}, {David}, {David},
  {de Laverny}, {De Luise}, {De March}, {de Martino}, {de Souza}, {de Torres},
  {Debosscher}, {del Pozo}, {Delbo}, {Delgado}, {Delgado}, {Di Matteo},
  {Diakite}, {Diener}, {Distefano}, {Dolding}, {Drazinos}, {Dur{\'a}n},
  {Edvardsson}, {Enke}, {Eriksson}, {Esquej}, {Eynard Bontemps}, {Fabre},
  {Fabrizio}, {Faigler}, {Falc{\~a}o}, {Farr{\`a}s Casas}, {Federici},
  {Fedorets}, {Fernique}, {Figueras}, {Filippi}, {Findeisen}, {Fonti},
  {Fraile}, {Fraser}, {Fr{\'e}zouls}, {Gai}, {Galleti}, {Garabato},
  {Garc{\'\i}a-Sedano}, {Garofalo}, {Garralda}, {Gavel}, {Gavras}, {Gerssen},
  {Geyer}, {Giacobbe}, {Gilmore}, {Girona}, {Giuffrida}, {Glass}, {Gomes},
  {Granvik}, {Gueguen}, {Guerrier}, {Guiraud}, {Guti{\'e}rrez-S{\'a}nchez},
  {Haigron}, {Hatzidimitriou}, {Hauser}, {Haywood}, {Heiter}, {Helmi}, {Heu},
  {Hilger}, {Hobbs}, {Hofmann}, {Holland}, {Huckle}, {Hypki}, {Icardi},
  {Jan{\ss}en}, {Jevardat de Fombelle}, {Jonker}, {Juh{\'a}sz}, {Julbe},
  {Karampelas}, {Kewley}, {Klar}, {Kochoska}, {Kohley}, {Kolenberg},
  {Kontizas}, {Kontizas}, {Koposov}, {Kordopatis}, {Kostrzewa-Rutkowska},
  {Koubsky}, {Lambert}, {Lanza}, {Lasne}, {Lavigne}, {Le Fustec}, {Le
  Poncin-Lafitte}, {Lebreton}, {Leccia}, {Leclerc}, {Lecoeur-Taibi},
  {Lenhardt}, {Leroux}, {Liao}, {Licata}, {Lindstr{\o}m}, {Lister}, {Livanou},
  {Lobel}, {L{\'o}pez}, {Managau}, {Mann}, {Mantelet}, {Marchal}, {Marchant},
  {Marconi}, {Marinoni}, {Marschalk{\'o}}, {Marshall}, {Martino}, {Marton},
  {Mary}, {Massari}, {Matijevi{\v{c}}}, {Mazeh}, {McMillan}, {Messina},
  {Michalik}, {Millar}, {Molina}, {Molinaro}, {Moln{\'a}r}, {Montegriffo},
  {Mor}, {Morbidelli}, {Morel}, {Morris}, {Mulone}, {Muraveva}, {Musella},
  {Nelemans}, {Nicastro}, {Noval}, {O'Mullane}, {Ord{\'e}novic},
  {Ord{\'o}{\~n}ez-Blanco}, {Osborne}, {Pagani}, {Pagano}, {Pailler},
  {Palacin}, {Palaversa}, {Panahi}, {Pawlak}, {Piersimoni}, {Pineau}, {Plachy},
  {Plum}, {Poggio}, {Poujoulet}, {Pr{\v{s}}a}, {Pulone}, {Racero}, {Ragaini},
  {Rambaux}, {Ramos-Lerate}, {Regibo}, {Reyl{\'e}}, {Riclet}, {Ripepi}, {Riva},
  {Rivard}, {Rixon}, {Roegiers}, {Roelens}, {Romero-G{\'o}mez}, {Rowell},
  {Royer}, {Ruiz-Dern}, {Sadowski}, {Sagrist{\`a} Sell{\'e}s}, {Sahlmann},
  {Salgado}, {Salguero}, {Sanna}, {Santana-Ros}, {Sarasso}, {Savietto},
  {Schultheis}, {Sciacca}, {Segol}, {Segovia}, {S{\'e}gransan}, {Shih},
  {Siltala}, {Silva}, {Smart}, {Smith}, {Solano}, {Solitro}, {Sordo}, {Soria
  Nieto}, {Souchay}, {Spagna}, {Spoto}, {Stampa}, {Steele},
  {Steidelm{\"u}ller}, {Stephenson}, {Stoev}, {Suess}, {Surdej}, {Szabados},
  {Szegedi-Elek}, {Tapiador}, {Taris}, {Tauran}, {Taylor}, {Teixeira},
  {Terrett}, {Teyssandier}, {Thuillot}, {Titarenko}, {Torra Clotet}, {Turon},
  {Ulla}, {Utrilla}, {Uzzi}, {Vaillant}, {Valentini}, {Valette}, {van Elteren},
  {Van Hemelryck}, {van Leeuwen}, {Vaschetto}, {Vecchiato}, {Veljanoski},
  {Viala}, {Vicente}, {Vogt}, {von Essen}, {Voss}, {Votruba}, {Voutsinas},
  {Walmsley}, {Weiler}, {Wertz}, {Wevers}, {Wyrzykowski}, {Yoldas},
  {{\v{Z}}erjal}, {Ziaeepour}, {Zorec}, {Zschocke}, {Zucker}, {Zurbach}, \&
  {Zwitter}}]{Gaia-Collaboration2018}
{Gaia Collaboration}, {Brown}, A.~G.~A., {Vallenari}, A., {et~al.} 2018, \aap,
  616, A1

\bibitem[{{Georgy} {et~al.}(2012){Georgy}, {Ekstr{\"o}m}, {Meynet}, {Massey},
  {Levesque}, {Hirschi}, {Eggenberger}, \& {Maeder}}]{Georgy+2012}
{Georgy}, C., {Ekstr{\"o}m}, S., {Meynet}, G., {et~al.} 2012, \aap, 542, A29

\bibitem[{{Gerke} {et~al.}(2015){Gerke}, {Kochanek}, \& {Stanek}}]{Gerke+2015}
{Gerke}, J.~R., {Kochanek}, C.~S., \& {Stanek}, K.~Z. 2015, \mnras, 450, 3289

\bibitem[{{Gilkis} {et~al.}(2016){Gilkis}, {Soker}, \& {Papish}}]{Gilkis+2016}
{Gilkis}, A., {Soker}, N., \& {Papish}, O. 2016, \apj, 826, 178

\bibitem[{{G{\"o}tberg} {et~al.}(2018){G{\"o}tberg}, {de Mink}, {Groh},
  {Kupfer}, {Crowther}, {Zapartas}, \& {Renzo}}]{Gotberg+2018}
{G{\"o}tberg}, Y., {de Mink}, S.~E., {Groh}, J.~H., {et~al.} 2018, \aap, 615,
  A78

\bibitem[{{Graur} {et~al.}(2017){Graur}, {Bianco}, {Modjaz}, {Shivvers},
  {Filippenko}, {Li}, \& {Smith}}]{Graur+2017}
{Graur}, O., {Bianco}, F.~B., {Modjaz}, M., {et~al.} 2017, \apj, 837, 121

\bibitem[{{Heger} {et~al.}(2003){Heger}, {Fryer}, {Woosley}, {Langer}, \&
  {Hartmann}}]{Heger+2003}
{Heger}, A., {Fryer}, C.~L., {Woosley}, S.~E., {Langer}, N., \& {Hartmann},
  D.~H. 2003, \apj, 591, 288

\bibitem[{{Heger} {et~al.}(2000){Heger}, {Langer}, \& {Woosley}}]{Heger+2000}
{Heger}, A., {Langer}, N., \& {Woosley}, S.~E. 2000, \apj, 528, 368

\bibitem[{Herant {et~al.}(1994)Herant, Benz, Hix, Fryer, \&
  Colgate}]{Herant+1994}
Herant, M., Benz, W., Hix, W.~R., Fryer, C.~L., \& Colgate, S.~A. 1994, The
  Astrophysical Journal, 435, 339

\bibitem[{{Higgins} {et~al.}(2021){Higgins}, {Sander}, {Vink}, \&
  {Hirschi}}]{Higgins+2021}
{Higgins}, E.~R., {Sander}, A.~A.~C., {Vink}, J.~S., \& {Hirschi}, R. 2021,
  \mnras, 505, 4874

\bibitem[{{Hirai} {et~al.}(2020){Hirai}, {Sato}, {Podsiadlowski},
  {Vigna-G{\'o}mez}, \& {Mandel}}]{Hirai+2020}
{Hirai}, R., {Sato}, T., {Podsiadlowski}, P., {Vigna-G{\'o}mez}, A., \&
  {Mandel}, I. 2020, \mnras, 499, 1154

\bibitem[{{Holt} {et~al.}(2019){Holt}, {Filippone}, \& {Pieper}}]{Holt+2019}
{Holt}, R.~J., {Filippone}, B.~W., \& {Pieper}, S.~C. 2019, \prc, 99, 055802

\bibitem[{{Ivanov} \& {Fern{\'a}ndez}(2021)}]{ivanov:21}
{Ivanov}, M. \& {Fern{\'a}ndez}, R. 2021, \apj, 911, 6

\bibitem[{Ivezi{\'{c}} {et~al.}(2019)Ivezi{\'{c}}, Kahn, Tyson, Abel, Acosta,
  Allsman, Alonso, AlSayyad, Anderson, Andrew, Angel, Angeli, Ansari,
  Antilogus, Araujo, Armstrong, Arndt, Astier, Aubourg, Auza, Axelrod, Bard,
  Barr, Barrau, Bartlett, Bauer, Bauman, Baumont, Bechtol, Bechtol, Becker,
  Becla, Beldica, Bellavia, Bianco, Biswas, Blanc, Blazek, Blandford, Bloom,
  Bogart, Bond, Booth, Borgland, Borne, Bosch, Boutigny, Brackett, Bradshaw,
  Brandt, Brown, Bullock, Burchat, Burke, Cagnoli, Calabrese, Callahan, Callen,
  Carlin, Carlson, Chandrasekharan, Charles-Emerson, Chesley, Cheu, Chiang,
  Chiang, Chirino, Chow, Ciardi, Claver, Cohen-Tanugi, Cockrum, Coles,
  Connolly, Cook, Cooray, Covey, Cribbs, Cui, Cutri, Daly, Daniel, Daruich,
  Daubard, Daues, Dawson, Delgado, Dellapenna, de~Peyster, de~Val-Borro, Digel,
  Doherty, Dubois, Dubois-Felsmann, Durech, Economou, Eifler, Eracleous,
  Emmons, Neto, Ferguson, Figueroa, Fisher-Levine, Focke, Foss, Frank, Freemon,
  Gangler, Gawiser, Geary, Gee, Geha, Gessner, Gibson, Gilmore, Glanzman,
  Glick, Goldina, Goldstein, Goodenow, Graham, Gressler, Gris, Guy, Guyonnet,
  Haller, Harris, Hascall, Haupt, Hernandez, Herrmann, Hileman, Hoblitt,
  Hodgson, Hogan, Howard, Huang, Huffer, Ingraham, Innes, Jacoby, Jain, Jammes,
  Jee, Jenness, Jernigan, Jevremovi{\'{c}}, Johns, Johnson, Johnson, Jones,
  Juramy-Gilles, Juri{\'{c}}, Kalirai, Kallivayalil, Kalmbach, Kantor, Karst,
  Kasliwal, Kelly, Kessler, Kinnison, Kirkby, Knox, Kotov, Krabbendam,
  Krughoff, Kub{\'{a}}nek, Kuczewski, Kulkarni, Ku, Kurita, Lage, Lambert,
  Lange, Langton, Guillou, Levine, Liang, Lim, Lintott, Long, Lopez, Lotz,
  Lupton, Lust, MacArthur, Mahabal, Mandelbaum, Markiewicz, Marsh, Marshall,
  Marshall, May, McKercher, McQueen, Meyers, Migliore, Miller, Mills, Miraval,
  Moeyens, Moolekamp, Monet, Moniez, Monkewitz, Montgomery, Morrison, Mueller,
  Muller, Arancibia, Neill, Newbry, Nief, Nomerotski, Nordby, O'Connor, Oliver,
  Olivier, Olsen, O'Mullane, Ortiz, Osier, Owen, Pain, Palecek, Parejko,
  Parsons, Pease, Peterson, Peterson, Petravick, Petrick, Petry, Pierfederici,
  Pietrowicz, Pike, Pinto, Plante, Plate, Plutchak, Price, Prouza, Radeka,
  Rajagopal, Rasmussen, Regnault, Reil, Reiss, Reuter, Ridgway, Riot, Ritz,
  Robinson, Roby, Roodman, Rosing, Roucelle, Rumore, Russo, Saha, Sassolas,
  Schalk, Schellart, Schindler, Schmidt, Schneider, Schneider, Schoening,
  Schumacher, Schwamb, Sebag, Selvy, Sembroski, Seppala, Serio, Serrano, Shaw,
  Shipsey, Sick, Silvestri, Slater, Smith, Smith, Sobhani, Soldahl,
  Storrie-Lombardi, Stover, Strauss, Street, Stubbs, Sullivan, Sweeney,
  Swinbank, Szalay, Takacs, Tether, Thaler, Thayer, Thomas, Thornton, Thukral,
  Tice, Trilling, Turri, Berg, Berk, Vetter, Virieux, Vucina, Wahl, Walkowicz,
  Walsh, Walter, Wang, Wang, Warner, Wiecha, Willman, Winters, Wittman, Wolff,
  Wood-Vasey, Wu, Xin, Yoachim, \& Zhan}]{ivezi:2019}
Ivezi{\'{c}}, {\v{Z}}., Kahn, S.~M., Tyson, J.~A., {et~al.} 2019, The
  Astrophysical Journal, 873, 111

\bibitem[{{Janka}(2012)}]{Janka2012}
{Janka}, H.-T. 2012, Annual Review of Nuclear and Particle Science, 62, 407

\bibitem[{{Janka}(2013)}]{Janka2013}
{Janka}, H.-T. 2013, \mnras, 434, 1355

\bibitem[{{Janka}(2017)}]{Janka+2017}
{Janka}, H.-T. 2017, \apj, 837, 84

\bibitem[{{Kroupa}(2001)}]{Kroupa2001}
{Kroupa}, P. 2001, \mnras, 322, 231

\bibitem[{{Kuroda} {et~al.}(2018){Kuroda}, {Kotake}, {Takiwaki}, \&
  {Thielemann}}]{Kuroda+2018}
{Kuroda}, T., {Kotake}, K., {Takiwaki}, T., \& {Thielemann}, F.-K. 2018,
  \mnras, 477, L80

\bibitem[{{Laplace} {et~al.}(2021){Laplace}, {Justham}, {Renzo}, {G{\"o}tberg},
  {Farmer}, {Vartanyan}, \& {de Mink}}]{Laplace+2021}
{Laplace}, E., {Justham}, S., {Renzo}, M., {et~al.} 2021, arXiv e-prints,
  arXiv:2102.05036

\bibitem[{{Limongi} \& {Chieffi}(2018)}]{Limongi+2018}
{Limongi}, M. \& {Chieffi}, A. 2018, \apjs, 237, 13

\bibitem[{{Lyman} {et~al.}(2016){Lyman}, {Bersier}, {James}, {Mazzali},
  {Eldridge}, {Fraser}, \& {Pian}}]{Lyman+2016}
{Lyman}, J.~D., {Bersier}, D., {James}, P.~A., {et~al.} 2016, \mnras, 457, 328

\bibitem[{{Mabanta} \& {Murphy}(2018)}]{Mabanta+Murphy2018}
{Mabanta}, Q.~A. \& {Murphy}, J.~W. 2018, \apj, 856, 22

\bibitem[{{Mabanta} {et~al.}(2019){Mabanta}, {Murphy}, \&
  {Dolence}}]{Mabanta+2019}
{Mabanta}, Q.~A., {Murphy}, J.~W., \& {Dolence}, J.~C. 2019, \apj, 887, 43

\bibitem[{{MacFadyen} \& {Woosley}(1999)}]{MacFadyen+Woosley1999}
{MacFadyen}, A.~I. \& {Woosley}, S.~E. 1999, \apj, 524, 262

\bibitem[{{Maeder} \& {Lequeux}(1982)}]{Maeder+1982}
{Maeder}, A. \& {Lequeux}, J. 1982, \aap, 114, 409

\bibitem[{{Mandel} \& {M{\"u}ller}(2020)}]{Mandel+2020}
{Mandel}, I. \& {M{\"u}ller}, B. 2020, \mnras, 499, 3214

\bibitem[{{Maund} {et~al.}(2004){Maund}, {Smartt}, {Kudritzki},
  {Podsiadlowski}, \& {Gilmore}}]{Maund+2004}
{Maund}, J.~R., {Smartt}, S.~J., {Kudritzki}, R.~P., {Podsiadlowski}, P., \&
  {Gilmore}, G.~F. 2004, \nat, 427, 129

\bibitem[{{Miller-Jones} {et~al.}(2021){Miller-Jones}, {Bahramian}, {Orosz},
  {Mandel}, {Gou}, {Maccarone}, {Neijssel}, {Zhao}, {Zi{\'o}{\l}kowski},
  {Reid}, {Uttley}, {Zheng}, {Byun}, {Dodson}, {Grinberg}, {Jung}, {Kim},
  {Marcote}, {Markoff}, {Rioja}, {Rushton}, {Russell}, {Sivakoff}, {Tetarenko},
  {Tudose}, \& {Wilms}}]{Miller-Jones+2021}
{Miller-Jones}, J. C.~A., {Bahramian}, A., {Orosz}, J.~A., {et~al.} 2021,
  Science, 371, 1046

\bibitem[{{Modjaz} {et~al.}(2016){Modjaz}, {Liu}, {Bianco}, \&
  {Graur}}]{Modjaz+2016}
{Modjaz}, M., {Liu}, Y.~Q., {Bianco}, F.~B., \& {Graur}, O. 2016, \apj, 832,
  108

\bibitem[{{M{\"u}ller}(2019)}]{Muller2019}
{M{\"u}ller}, B. 2019, Annual Review of Nuclear and Particle Science, 69, 253

\bibitem[{{Nadezhin}(1980)}]{Nadezhin1980}
{Nadezhin}, D.~K. 1980, \apss, 69, 115

\bibitem[{{Neijssel} {et~al.}(2021){Neijssel}, {Vinciguerra},
  {Vigna-G{\'o}mez}, {Hirai}, {Miller-Jones}, {Bahramian}, {Maccarone}, \&
  {Mandel}}]{Neijssel+2021}
{Neijssel}, C.~J., {Vinciguerra}, S., {Vigna-G{\'o}mez}, A., {et~al.} 2021,
  \apj, 908, 118

\bibitem[{{Neustadt} {et~al.}(2021){Neustadt}, {Kochanek}, {Stanek},
  {Basinger}, {Jayasinghe}, {Garling}, {Adams}, \& {Gerke}}]{Neustadt+2021}
{Neustadt}, J.~M.~M., {Kochanek}, C.~S., {Stanek}, K.~Z., {et~al.} 2021,
  \mnras, 508, 516

\bibitem[{{Nugis} \& {Lamers}(2000)}]{Nugis+2000}
{Nugis}, T. \& {Lamers}, H.~J.~G.~L.~M. 2000, \aap, 360, 227

\bibitem[{{O'Connor} \& {Ott}(2011)}]{OConnor+2011}
{O'Connor}, E. \& {Ott}, C.~D. 2011, \apj, 730, 70

\bibitem[{{Ott} {et~al.}(2018){Ott}, {Roberts}, {da Silva Schneider}, {Fedrow},
  {Haas}, \& {Schnetter}}]{Ott+2018}
{Ott}, C.~D., {Roberts}, L.~F., {da Silva Schneider}, A., {et~al.} 2018, \apjl,
  855, L3

\bibitem[{{Patton} \& {Sukhbold}(2020)}]{Patton+2020}
{Patton}, R.~A. \& {Sukhbold}, T. 2020, \mnras, 499, 2803

\bibitem[{{Patton} {et~al.}(2021){Patton}, {Sukhbold}, \&
  {Eldridge}}]{Patton+2021}
{Patton}, R.~A., {Sukhbold}, T., \& {Eldridge}, J.~J. 2021, arXiv e-prints,
  arXiv:2106.05978

\bibitem[{{Paxton} {et~al.}(2011){Paxton}, {Bildsten}, {Dotter}, {Herwig},
  {Lesaffre}, \& {Timmes}}]{Paxton+2011}
{Paxton}, B., {Bildsten}, L., {Dotter}, A., {et~al.} 2011, \apjs, 192, 3

\bibitem[{{Paxton} {et~al.}(2013){Paxton}, {Cantiello}, {Arras}, {Bildsten},
  {Brown}, {Dotter}, {Mankovich}, {Montgomery}, {Stello}, {Timmes}, \&
  {Townsend}}]{Paxton+2013}
{Paxton}, B., {Cantiello}, M., {Arras}, P., {et~al.} 2013, \apjs, 208, 4

\bibitem[{{Paxton} {et~al.}(2015){Paxton}, {Marchant}, {Schwab}, {Bauer},
  {Bildsten}, {Cantiello}, {Dessart}, {Farmer}, {Hu}, {Langer}, {Townsend},
  {Townsley}, \& {Timmes}}]{Paxton+2015}
{Paxton}, B., {Marchant}, P., {Schwab}, J., {et~al.} 2015, \apjs, 220, 15

\bibitem[{{Paxton} {et~al.}(2018){Paxton}, {Schwab}, {Bauer}, {Bildsten},
  {Blinnikov}, {Duffell}, {Farmer}, {Goldberg}, {Marchant}, {Sorokina},
  {Thoul}, {Townsend}, \& {Timmes}}]{Paxton+2018}
{Paxton}, B., {Schwab}, J., {Bauer}, E.~B., {et~al.} 2018, \apjs, 234, 34

\bibitem[{{Paxton} {et~al.}(2019){Paxton}, {Smolec}, {Schwab}, {Gautschy},
  {Bildsten}, {Cantiello}, {Dotter}, {Farmer}, {Goldberg}, {Jermyn}, {Kanbur},
  {Marchant}, {Thoul}, {Townsend}, {Wolf}, {Zhang}, \& {Timmes}}]{Paxton+2019}
{Paxton}, B., {Smolec}, R., {Schwab}, J., {et~al.} 2019, \apjs, 243, 10

\bibitem[{{Perley} {et~al.}(2020){Perley}, {Fremling}, {Sollerman}, {Miller},
  {Dahiwale}, {Sharma}, {Bellm}, {Biswas}, {Brink}, {Bruch}, {De}, {Dekany},
  {Drake}, {Duev}, {Filippenko}, {Gal-Yam}, {Goobar}, {Graham}, {Graham}, {Ho},
  {Irani}, {Kasliwal}, {Kim}, {Kulkarni}, {Mahabal}, {Masci}, {Modak}, {Neill},
  {Nordin}, {Riddle}, {Soumagnac}, {Strotjohann}, {Schulze}, {Taggart},
  {Tzanidakis}, {Walters}, \& {Yan}}]{Perley+2020}
{Perley}, D.~A., {Fremling}, C., {Sollerman}, J., {et~al.} 2020, \apj, 904, 35

\bibitem[{{Podsiadlowski} {et~al.}(1992){Podsiadlowski}, {Joss}, \&
  {Hsu}}]{Podsiadlowski+1992}
{Podsiadlowski}, P., {Joss}, P.~C., \& {Hsu}, J.~J.~L. 1992, \apj, 391, 246

\bibitem[{{Powell} \& {M{\"u}ller}(2020)}]{Powell+2020}
{Powell}, J. \& {M{\"u}ller}, B. 2020, \mnras, 494, 4665

\bibitem[{{Quataert} {et~al.}(2019){Quataert}, {Lecoanet}, \&
  {Coughlin}}]{quataert:19}
{Quataert}, E., {Lecoanet}, D., \& {Coughlin}, E.~R. 2019, \mnras, 485, L83

\bibitem[{{Quataert} \& {Shiode}(2012)}]{Quataert+Shiode2012}
{Quataert}, E. \& {Shiode}, J. 2012, \mnras, 423, L92

\bibitem[{{Renzo} {et~al.}(2017){Renzo}, {Ott}, {Shore}, \& {de
  Mink}}]{Renzo+2017}
{Renzo}, M., {Ott}, C.~D., {Shore}, S.~N., \& {de Mink}, S.~E. 2017, \aap, 603,
  A118

\bibitem[{{Rom{\'a}n-Garza} {et~al.}(2021){Rom{\'a}n-Garza}, {Bavera},
  {Fragos}, {Zapartas}, {Misra}, {Andrews}, {Coughlin}, {Dotter}, {Kovlakas},
  {Serra}, {Qin}, {Rocha}, \& {Tran}}]{Roman-Garza+2021}
{Rom{\'a}n-Garza}, J., {Bavera}, S.~S., {Fragos}, T., {et~al.} 2021, \apjl,
  912, L23

\bibitem[{{Ryder} {et~al.}(2018){Ryder}, {Van Dyk}, {Fox}, {Zapartas}, {de
  Mink}, {Smith}, {Brunsden}, {Azalee Bostroem}, {Filippenko}, {Shivvers}, \&
  {Zheng}}]{Ryder+2018}
{Ryder}, S.~D., {Van Dyk}, S.~D., {Fox}, O.~D., {et~al.} 2018, \apj, 856, 83

\bibitem[{{Sander} {et~al.}(2019){Sander}, {Hamann}, {Todt}, {Hainich},
  {Shenar}, {Ramachandran}, \& {Oskinova}}]{Sander+2019}
{Sander}, A.~A.~C., {Hamann}, W.~R., {Todt}, H., {et~al.} 2019, \aap, 621, A92

\bibitem[{{Schneider} {et~al.}(2021){Schneider}, {Podsiadlowski}, \&
  {M{\"u}ller}}]{Schneider+2021}
{Schneider}, F.~R.~N., {Podsiadlowski}, P., \& {M{\"u}ller}, B. 2021, \aap,
  645, A5

\bibitem[{{Shivvers} {et~al.}(2019){Shivvers}, {Filippenko}, {Silverman},
  {Zheng}, {Foley}, {Chornock}, {Barth}, {Cenko}, {Clubb}, {Fox},
  {Ganeshalingam}, {Graham}, {Kelly}, {Kleiser}, {Leonard}, {Li}, {Matheson},
  {Mauerhan}, {Modjaz}, {Serduke}, {Shields}, {Steele}, {Swift}, {Wong}, \&
  {Yuk}}]{Shivvers+2019}
{Shivvers}, I., {Filippenko}, A.~V., {Silverman}, J.~M., {et~al.} 2019, \mnras,
  482, 1545

\bibitem[{{Smith}(2014)}]{Smith2014}
{Smith}, N. 2014, \araa, 52, 487

\bibitem[{{Smith} {et~al.}(2011){Smith}, {Li}, {Filippenko}, \&
  {Chornock}}]{Smith+2011}
{Smith}, N., {Li}, W., {Filippenko}, A.~V., \& {Chornock}, R. 2011, \mnras,
  412, 1522

\bibitem[{{Soker}(2019)}]{Soker2019}
{Soker}, N. 2019, Research in Astronomy and Astrophysics, 19, 095

\bibitem[{{Spera} {et~al.}(2015){Spera}, {Mapelli}, \& {Bressan}}]{Spera+2015}
{Spera}, M., {Mapelli}, M., \& {Bressan}, A. 2015, \mnras, 451, 4086

\bibitem[{{Spruit}(2002)}]{Spruit2002}
{Spruit}, H.~C. 2002, \aap, 381, 923

\bibitem[{{Sravan} {et~al.}(2019){Sravan}, {Marchant}, \&
  {Kalogera}}]{Sravan+2019}
{Sravan}, N., {Marchant}, P., \& {Kalogera}, V. 2019, \apj, 885, 130

\bibitem[{{Sravan} {et~al.}(2018){Sravan}, {Marchant}, {Kalogera}, \&
  {Margutti}}]{Sravan+2018}
{Sravan}, N., {Marchant}, P., {Kalogera}, V., \& {Margutti}, R. 2018, \apjl,
  852, L17

\bibitem[{{Sravan} {et~al.}(2020){Sravan}, {Marchant}, {Kalogera},
  {Milisavljevic}, \& {Margutti}}]{Sravan+2020}
{Sravan}, N., {Marchant}, P., {Kalogera}, V., {Milisavljevic}, D., \&
  {Margutti}, R. 2020, \apj, 903, 70

\bibitem[{{Sukhbold} {et~al.}(2016){Sukhbold}, {Ertl}, {Woosley}, {Brown}, \&
  {Janka}}]{Sukhbold+2016}
{Sukhbold}, T., {Ertl}, T., {Woosley}, S.~E., {Brown}, J.~M., \& {Janka}, H.-T.
  2016, \apj, 821, 38

\bibitem[{{Sukhbold} \& {Woosley}(2014)}]{Sukhbold+2014}
{Sukhbold}, T. \& {Woosley}, S.~E. 2014, \apj, 783, 10

\bibitem[{{Tur} {et~al.}(2007){Tur}, {Heger}, \& {Austin}}]{Tur+2007}
{Tur}, C., {Heger}, A., \& {Austin}, S.~M. 2007, \apj, 671, 821

\bibitem[{{Van Dyk} {et~al.}(2011){Van Dyk}, {Li}, {Cenko}, {Kasliwal},
  {Horesh}, {Ofek}, {Kraus}, {Silverman}, {Arcavi}, {Filippenko}, {Gal-Yam},
  {Quimby}, {Kulkarni}, {Yaron}, \& {Polishook}}]{van-Dyk+2011}
{Van Dyk}, S.~D., {Li}, W., {Cenko}, S.~B., {et~al.} 2011, \apjl, 741, L28

\bibitem[{{Van Dyk} {et~al.}(2019){Van Dyk}, {Zheng}, {Maund}, {Brink},
  {Srinivasan}, {Andrews}, {Smith}, {Leonard}, {Morozova}, {Filippenko},
  {Conner}, {Milisavljevic}, {de Jaeger}, {Long}, {Isaacson}, {Crossfield},
  {Kosiarek}, {Howard}, {Fox}, {Kelly}, {Piro}, {Littlefair}, {Dhillon},
  {Wilson}, {Butterley}, {Yunus}, {Channa}, {Jeffers}, {Falcon}, {Ross},
  {Hestenes}, {Stegman}, {Zhang}, \& {Kumar}}]{van-Dyk+2019}
{Van Dyk}, S.~D., {Zheng}, W., {Maund}, J.~R., {et~al.} 2019, \apj, 875, 136

\bibitem[{{van Loon}(2006)}]{van-Loon2006}
{van Loon}, J.~T. 2006, in Astronomical Society of the Pacific Conference
  Series, Vol. 353, Stellar Evolution at Low Metallicity: Mass Loss,
  Explosions, Cosmology, ed. {H.~J.~G.~L.~M.~Lamers, N.~Langer, T.~Nugis \&
  K.~Annuk}, 211

\bibitem[{{Vartanyan} {et~al.}(2019){Vartanyan}, {Burrows}, \&
  {Radice}}]{Vartanyan+2019}
{Vartanyan}, D., {Burrows}, A., \& {Radice}, D. 2019, \mnras, 489, 2227

\bibitem[{{Vartanyan} {et~al.}(2021){Vartanyan}, {Laplace}, {Renzo},
  {G{\"o}tberg}, {Burrows}, \& {de Mink}}]{Vartanyan+2021}
{Vartanyan}, D., {Laplace}, E., {Renzo}, M., {et~al.} 2021, \apjl, 916, L5

\bibitem[{{Vink} {et~al.}(2000){Vink}, {de Koter}, \& {Lamers}}]{Vink+2000}
{Vink}, J.~S., {de Koter}, A., \& {Lamers}, H.~J.~G.~L.~M. 2000, \aap, 362, 295

\bibitem[{{Vink} {et~al.}(2001){Vink}, {de Koter}, \& {Lamers}}]{Vink+2001}
{Vink}, J.~S., {de Koter}, A., \& {Lamers}, H.~J.~G.~L.~M. 2001, \aap, 369, 574

\bibitem[{{Yoon} {et~al.}(2017){Yoon}, {Dessart}, \& {Clocchiatti}}]{Yoon+2017}
{Yoon}, S.-C., {Dessart}, L., \& {Clocchiatti}, A. 2017, \apj, 840, 10

\bibitem[{{Yoon} {et~al.}(2012){Yoon}, {Dierks}, \& {Langer}}]{Yoon+2012}
{Yoon}, S.-C., {Dierks}, A., \& {Langer}, N. 2012, \aap, 542, A113

\bibitem[{{Yoon} {et~al.}(2010){Yoon}, {Woosley}, \& {Langer}}]{Yoon+2010}
{Yoon}, S.-C., {Woosley}, S.~E., \& {Langer}, N. 2010, \apj, 725, 940

\bibitem[{{Zapartas} {et~al.}(2017){Zapartas}, {de Mink}, {Van Dyk}, {Fox},
  {Smith}, {Bostroem}, {de Koter}, {Filippenko}, {Izzard}, {Kelly}, {Neijssel},
  {Renzo}, \& {Ryder}}]{Zapartas+2017b}
{Zapartas}, E., {de Mink}, S.~E., {Van Dyk}, S.~D., {et~al.} 2017, \apj, 842,
  125

\end{thebibliography}

\end{document}